\documentclass{svjour3} 
\usepackage{booktabs} 

\usepackage{amsmath}
\usepackage{siunitx}
\usepackage{amssymb}
\usepackage{booktabs}
\usepackage{verbatim} 
\usepackage{subfigure}
\usepackage{balance}
\usepackage{graphicx}
\usepackage{tikz}
\usepackage{algpseudocode}
\usepackage[bottom]{footmisc}
\usetikzlibrary{fit,positioning}
\usepackage[utf8]{inputenc}
\usepackage[english]{babel}
\usepackage{caption}
\usepackage{mathtools}
\usepackage[hyphens]{url}
\usepackage[bookmarks=false]{hyperref}
\usepackage[numbers]{natbib}

\usepackage{xcolor}

\newcommand{\lanyu}[1]{\textcolor{black}{#1}}
\newcommand{\lanyus}[1]{\textcolor{black}{#1}}
\newcommand{\dz}[1]{\textcolor{black}{#1}}
\newcommand{\lanyurv}[1]{\textcolor{black}{#1}}
\newcommand{\lanyusrv}[1]{\textcolor{black}{#1}}
\newcommand{\yz}[1]{\textcolor{black}{#1}}
\newcommand{\shang}[1]{\textcolor{black}{#1}}

\begin{document}

\title{FauxWard: A Graph Neural Network Approach to Fauxtography Detection Using Social Media Comments}
\author{Lanyu Shang \and Yang Zhang \and Daniel Zhang  \and Dong Wang}
\institute{Lanyu Shang \and Yang Zhang \and Daniel Zhang  \and Dong Wang \at Department of Computer Science and Engineering\\
University of Notre Dame, Notre Dame, IN 46556\\
\email{\{lshang, yzhang42, yzhang40, dwang5\}@nd.edu}}

\maketitle

\begin{abstract}
Online social media has been a popular source for people to consume and share news content. More recently, the spread of misinformation online has caused widespread concerns. In this work, we focus on a critical task of detecting fauxtography on social media where the image and associated text together convey misleading information. Many efforts have been made to mitigate misinformation online, but we found that the fauxtography problem has not been fully addressed by existing work. 
\shang{Solutions focusing on detecting fake images or misinformed texts alone on social media often fail to identify the misinformation delivered together by the image and the associated text of a fauxtography post.}
\yz{In this paper, we develop \shang{FauxWard}, a novel graph convolutional neural network framework that explicitly explores the \shang{complex} information extracted from a user comment network of a social media post  to effectively identify fauxtography.} FauxWard is content-free in the sense that it does not analyze the visual or textual contents of the post itself, which makes it robust against sophisticated fauxtography uploaders who intentionally craft image-centric posts by editing either the text or image content. We evaluate FauxWard on two real-world datasets collected from mainstream social media platforms (i.e., Reddit and Twitter). The results show that FauxWard is both effective and efficient in identifying fauxtography posts on social media.
\end{abstract}



\section{Introduction}
\label{sec:intro}

In recent years, social media has become a popular channel for people to consume and share news content~\cite{kwak2010twitter,wang2018age}. However, the spread of misinformation on social media platforms has raised many concerns, and a significant amount of efforts have been made to reduce the diffusion of misinformation online~\cite{zhou2018fake,shang2019vulnercheck}. For example, leading social media platforms (e.g., Facebook and Google) have stepped up to tackle and prevent the spread of fake news~\cite{fakenews}. Many solutions have been developed to combat misinformation propagation on online social media, including the analysis of news content~\cite{volkova2017separating}, the assessment of news source credibility~\cite{zhang2018fake}, and a set of fact-checking techniques~\cite{vo2018rise}.
In this paper, we focus on an important but largely unsolved problem of detecting ``fauxtography" where the image(s) and the associated text of a social media post conveys a questionable or outright false sense of the events it seems to depict \cite{cooper2007concise}.

\lanyu{The increasing popularity of visual content on online social media~\cite{photo,shang2019towards} has motivated our study of detecting fauxtography. For example, photos have been recognized as the primary type of content on social media. A social media post accompanied by an image is ten times more likely to attract engagements (e.g., click, like, or share)~\cite{engagement}. In particular, on Twitter, tweets with image content could attract} 18\% more clicks, 89\% more likes, and 150\% more retweets than tweets without images~\cite{twitterengagement}.

\begin{figure}[!htb]
    \begin{center}
     \subfigure[][Fake Image, False Text]{
         \centering
         \includegraphics[width=0.4\linewidth, height = 3cm]{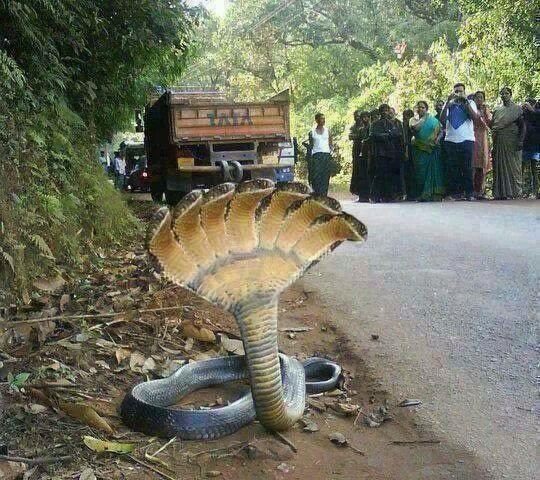}
         \label{fig:fa}
     }
     \subfigure[][Real Image, False Text]{
         \includegraphics[width=0.4\linewidth, height = 3cm]{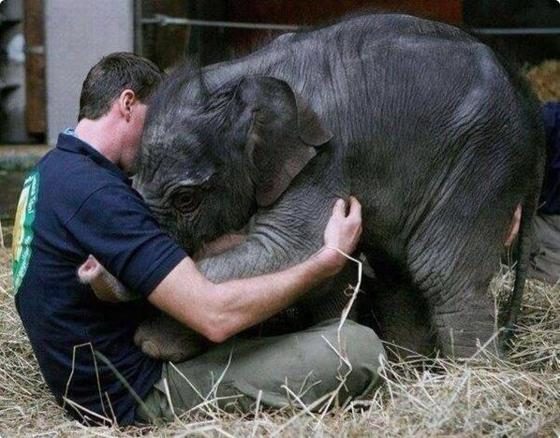}
        \label{fig:fb}
     }
     \subfigure[][Fake Image, True Text]{
         \includegraphics[width=0.4\linewidth, height = 3cm]{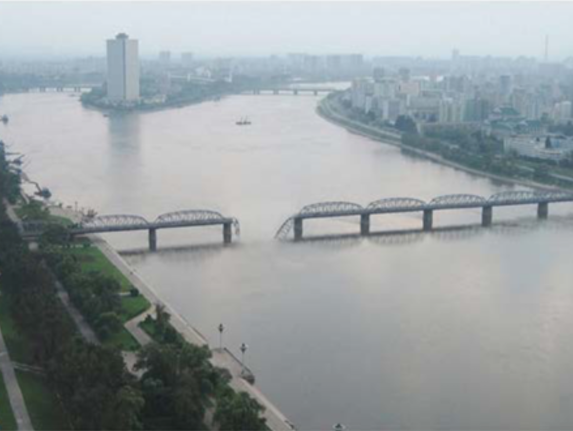}
        \label{fig:fd}
     }
     \subfigure[][Real Image, True Text]{
         \includegraphics[width=0.4\linewidth, height = 3cm]{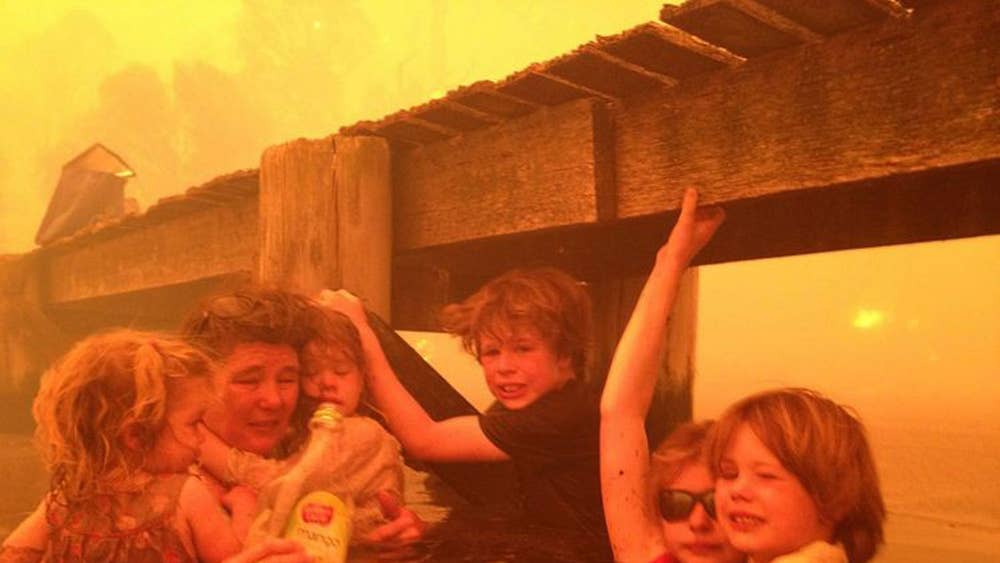}
        \label{fig:fc}
     }
     \end{center}
     {\small
     \lanyu{\textbf{Image (a)} was titled ``A rare seven-headed snake found in Honduras."}\\
     \lanyu{\textbf{Image (b)} was titled ``A baby elephant lost her mother to poachers."}\\
     \lanyu{\textbf{Image (c)} was titled ``At least one person was killed after part of a bridge in China's Zhengzhou city collapsed."}\\
    \lanyu{\textbf{Image (d)} was titled ``Team UMA (Utsav Melbourne Association) is looking for your support and donations towards helping fellow Aussies who have lost everything in the prevailing bushfires."}}
     \caption{Examples of Fauxtography on Social Media}
     \label{fig:fake}
     \vspace{-0.2in}
 \end{figure}

With the growing presence of image-centric content, social media has become a rich playground for the propagation of misinformation \cite{wang2015social,wang2019age,rashid2020covidsens,zhang2019social}. For example, fake images about sightings of creepy killer clowns have caused national hysteria in 2016 in the USA\footnote{https://www.theverge.com/2016/10/7/13191788/clown-attack-threats-2016-panic-hoax-debunked}. 
\lanyus{In this paper, we investigate a unique type of misinformation on social media, namely fauxtography, where an image and its context (e.g., the associated text of the image-centric post) jointly convey misleading information to the viewers of the content.}
For example, all images in Figure~\ref{fig:fake} fall under our definition of fauxtography. In particular,  the text of image (a) claims that \lanyu{a seven-headed snake was found in Honduras}, while in fact, the image was manipulated (i.e., photoshopped) and the claim itself was false. 
Image~(b) claims a baby elephant lost her mother to poachers. While the image itself was a genuine photo (i.e., unedited), the baby elephant did not lose her mother to poachers and the photo was taken when the baby elephant was playing with her keeper at Munich zoo.
Image (c) claims at least one person was killed after part of a bridge in China's Zhengzhou city collapsed. Although the claim itself is a true event\footnote{http://www.globaltimes.cn/content/759679.shtml}, the image is misleading because the collapsed bridge in the image was intentionally manipulated (i.e., photoshopped) to deliver a false sense that the collapsed bridge was a huge bridge crossing a wide river while the actual one was an overpass in the city.
Last, image (d) accompanies a donation post for refugees during the recent bushfire in Australia.  While both the image and text are real, it is misleading because the image was taken from an earlier Australian bushfire in 2013\footnote{https://www.independent.co.uk/news/world/australasia/family-took-refuge-in-a-lake-to-escape-the-aussie-bushfires-8444881.html} and was used to exaggerate the severity of the fire.
In short, all the above cases will be considered as fauxtography because the images and the associated texts together convey misleading information.

\shang{The nature of the fauxtography detection problem requires a joint consideration of not only the truthfulness of the image or its associated text, but also the relation between them. 
Any content-based method that asserts the veracity of post content (e.g., the image or the text of the post) will be insufficient to address this problem. 
For example, the ``image forgery detection" solutions were developed to detect image manipulation, such as copy-and-move~\cite{fridrich2003detection}, splicing~\cite{huynh2016robust}, and image-retouch~\cite{zhang2020pqa}.
However, they only focus on detecting the verity of an image (i.e., fake image) but ignore the necessary context (e.g., the associated text) in an image-centric post. 
Hence, such kind of solutions cannot be directly applied to solve the fauxtography problem. For example, we observe that real images can convey misleading information that cannot be easily detected (e.g., images (b) and (d) in Figure \ref{fig:fake}).  
Furthermore, fact-checking solutions that are focusing on inferring the truthfulness of textual claims on social media~\cite{vo2018rise,R33} are also insufficient to fully address the fauxtography problem, especially when the associated text is true but the image is fake (e.g., images (c) in Figure \ref{fig:fake}). 
More recently, a few fake news detection solutions were proposed to leverage both the visual features extracted from images and text information in news articles to identify fake news~\cite{yang2018ti}. However, it is also insufficient to address the fauxtography problem where a falsified association between a real image and true text together convey misleading information (e.g., image (d) in Figure \ref{fig:fake}). 
Therefore, it is very difficult for these content-based solutions to effectively detect fauxtography.
}

\shang{In this paper, we develop FauxWard, a novel graph convolutional neural network based approach that can effectively track down fauxtography posts on online social media.
To overcome the limitation of content-based solutions that can be misled by posts with real image and/or true text, the proposed FauxWard framework is content-free in that it approaches the fauxtography detection problem without analyzing the content (i.e., both text and image) of the post. The content-free nature of FauxWard makes it robust against sophisticated content crafters who can intentionally modify the presentation and the description of the images~\cite{zhang2018streamguard,zhang2018crowdsourcing}. In particular, it leverages the user comments of the corresponding post and learns valuable information (e.g., textual content and replying pattern) from the comments to identify the fauxtography post.
For example, social media users often discuss more on the image verity of the fauxtography posts, and comments that directly debunk the fauxtography post usually receive more endorsements from other users. In contrast, topics discussed in the comments of non-fauxtography posts appear to be more diversified and users tend to have less debunk and endorsement behavior in their comments for non-fauxtography. 
Current solutions leveraging user comments only focus on the textual contents but ignore the replying pattern of user comments~\cite{qian2018neural,cui2019same}. Previous work~\cite{zhang2018fauxbuster} adopted the random walk based algorithms to extract the topological features of the user comment network. We observe that the topological feature of the comment structure identified by such an approach is often insufficient and over-simplified, which leads to suboptimal performance in detecting fauxtography posts with complex user comment networks.
In FauxWard, we develop a principled framework to extract a diversified set of valuable features (e.g., linguistic features, semantic features, and metadata features) from user comments to systematically characterize fauxtography. FauxWard then aggregates the extracted comment features from the user comment network of various sizes and structures through a graph convolutional neural network framework to track down fauxtography effectively.}

\yz{To the best of our knowledge, the Fauxward is the first graph neural network based approach to address the fauxtography detection problem on online social media.}
\lanyurv{The graph convolutional neural network design allows FauxWard to effectively learn graph-level representations of the user comment networks that vary in sizes and topological structures.}
We evaluate the performance of FauxWard on two real-world datasets collected from two mainstream social media platforms, Reddit and Twitter. The results show that our scheme significantly outperforms state-of-the-art fauxtography detection baselines in terms of both detection effectiveness and efficiency.

\lanyu{A preliminary version of this work has been published in \cite{zhang2018fauxbuster} to investigate the \textit{fauxtography detection} problem on online social media. This paper is a significant extension of the previous work (i.e., FauxBuster) in the following aspects. First, we identified a new challenge in effectively capturing the underlying topological structure of the user comment network where the size and the structure of the network differ in each post. We re-formulate the fauxtography detection problem under this new challenge.
Second, we developed a new graph convolutional neural network approach, FauxWard, to address the above challenge by modeling the fauxtography detection task as a graph classification problem and jointly leveraging the linguistic and semantic attributes of the comments and the topological characteristics of the user comment network to identify fauxtography posts (Section \ref{sec:solution}). 
Third, we collected two new datasets from Reddit and Twitter that include more recent fauxtography posts (until 2019) to evaluate the performance and robustness of the proposed scheme in a more realistic scenario (Section \ref{sec:data}). Fourth, we compared the FauxWard scheme with two additional state-of-the-art baselines on fauxtography and fake news detection to comprehensively study the effectiveness and efficiency of all compared schemes~(Section \ref{sec:eval}). Fifth, we extended the related work by reviewing recent works on graph neural networks (Section \ref{sec:related}). 
}

\section{Related Work}
\label{sec:related}

\subsection{Fauxtography}
\lanyus{The phenomenon of ``Fauxtography" first appeared in the 2006 Lebanon War when digitally manipulated photographs were used in news articles~\cite{cooper2007concise}. Cooper \emph{et al.} defined fauxtography as}
``visual images, especially news photographs, which convey a questionable (or outright false) sense of the events they seem to depict"~\cite{cooper2007concise}. Examples of fauxtography include taking photos of a staged event, using images from another irrelevant event, using digital editing tools (e.g., Photoshop) to manipulate the image, \lanyu{applying} special photography technique (e.g., wide-angle close-ups) to take \lanyu{photos} to exaggerate the event\lanyu{, and generating fake images with advanced computer vision technology (e.g., Deepfake)}. The fauxtography phenomenon has also been observed in social science, but no practical solution has been developed~\cite{yao2017shaping,gupta2013faking}. \lanyurv{In this paper, we develop the FauxWard, a graph neural network approach dedicated to addressing the fauxtography detection problem on online social media.}

\subsection{Image Forgery Detection}
Image forgery is closely related to \lanyu{the fauxtography detection} problem. 
\lanyus{A significant amount of efforts have been made to address the image forgery problem.}
For example, Huynh-Kha \emph{et al.} 
\lanyus{proposed an algorithm to detect image forgery where images are manipulated}
by copy-move, splicing, or both in the same image~\cite{huynh2016robust}.
\lanyu{Pun \emph{et al.} proposed a segmentation-based framework to identify image copy-move  forgery~\cite{pun2015image}.}
\lanyus{Bayar \emph{et al.} developed a convolutional neural networks based framework to suppress image contents and automatically detect image manipulations~\cite{bayar2016deep}.}
\lanyu{Matern \emph{et al.} proposed a gradient-based scheme to detect image forgery by validating the consistency of illumination between pairs of objects on the image~\cite{matern2019gradient}.}
Gupta \emph{et al.} characterized the phenomenon of fake image propagation on Twitter during a disaster event and developed a supervised detection scheme~\cite{gupta2013faking}. However, these schemes only focus on the visual content of the images while ignoring the associated context (e.g., text).  Therefore, they cannot address the fauxtography problem when the uploaders leverage real images to convey misleading information. In contrast, FauxWard assumes the fauxtography detection must consider both images and their contexts under a holistic analytical framework. 

\subsection{Misinformation Detection}
\lanyus{Misinformation has emerged as a critical issue on online social media and several solutions have been developed to mitigate the spread of misinformation~\cite{zhang2018towards,wang2014using,wang2013recursive,wang2013exploitation,wang2015reliable}. For example,} 
Yin \textit{et al.} proposed the first fact-checking scheme~\emph{Truth Finder} that uses a Bayesian-based heuristic algorithm to combat misinformation 
\lanyus{by finding true facts from a large amount of conflicting information~\cite{R33}.}
Wang \emph{et al.} developed an estimation-maximization algorithm that identifies truthful online social media posts by explicitly considering the reliability of data sources~\cite{R1}. Zhang \textit{et al.} developed a dynamic truth discovery model to incorporate physical constraints and temporal dependencies into the detection of evolving truth~\cite{zhang2017constraint}. Vo \textit{et al.} developed a fake news detection scheme that leverages the users who actively debunk fake information on social media, and recommends fact-checking URLs posted from these users~\cite{vo2018rise}.
\lanyu{P{\'e}rez-Rosas \textit{et al.} proposed a natural language processing based scheme to automatically identify fake content in online news media~\cite{perez2017automatic}.}
\shang{Yang \textit{et al.} developed a convolutional neural network framework to detect fake news by leveraging textual and visual features extracted from news articles~\cite{yang2018ti}.}
\shang{However, these content-based solutions cannot fully address the fauxtography problem and are insufficient to capture sophisticated fauxtography posts that covey misinformation using real images and true texts.}
\lanyus{In contrast, FauxWard leverages the ``wisdom of the crowd" and explores useful clues in the user comments to effectively identify misinformation of image-centric posts on social media.}

\subsection{Graph Neural Network}
\lanyu{Our work is related to Graph Neural Network (GNN)~\cite{zhou2018graph,wu2019comprehensive}. GNN is a deep learning based method that can be applied to complex graph-structured data in the non-Euclidean domain, including social networks, protein-protein interaction networks, and knowledge graphs~\cite{zhang2020multi,hamilton2017inductive,fout2017protein,hamaguchi2017knowledge}. 
For example, Ying \emph{et al.} developed a random walk based graph convolutional network solution to generate high-quality recommendations in large scale recommender systems~\cite{ying2018graph}. \lanyusrv{Li \emph{et al.} proposed an adaptive GNN framework that predicts toxicological effects of chemical compounds by taking arbitrary graph-structured molecular data as input}~\cite{li2018adaptive}.
Schlichtkrull~\emph{et~al.} developed a relational GNN scheme to effectively model the multi-relational data in knowledge bases~\cite{schlichtkrull2018modeling}.} \lanyusrv{Chen \emph{et al.} proposed a batched training scheme to classify research topics on citation networks by efficiently training GNN models on large and dense graphs~\cite{chen2018fastgcn}. Nguyen \emph{et al.} developed an argument-aware graph convolutional neural network model to detect events of interest in news articles~\cite{nguyen2018graph}. Current GNN-based approaches often assume a homogeneous set of nodes in the input graph and ignores the complex information embedded in the nodes.}
\yz{In this paper, we propose a graph convolutional neural network framework that \lanyusrv{leverages key features captured from user comments and effectively classify}
user comment networks of various sizes and structures with a cluster-based pooling strategy.}
\lanyusrv{To the best of our knowledge, FauxWard is the first GNN-based approach to detect fauxtography on social media. }

\newtheorem{myDef}{Definition}
\section{Problem Statement}
\label{sec:problem}

In this section, we present the fauxtography detection problem on online social media. We first define a few key terms that will be used in the problem formulation.

\begin{figure}[!htb]
 \centering
     \subfigure[][Claim and Image]{
         \includegraphics[width=0.48\textwidth, height = 5cm]{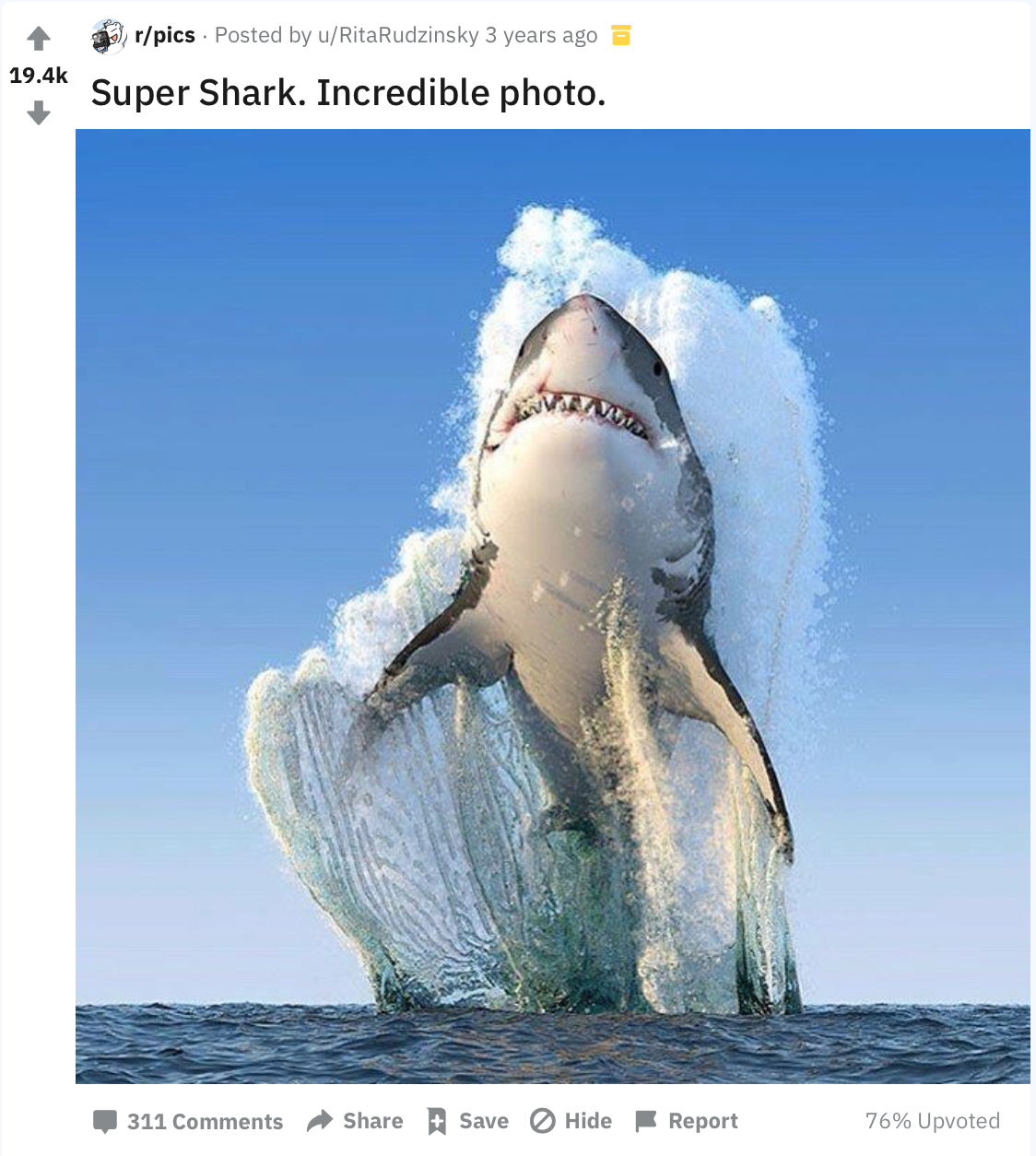}
         \label{fig:examplepost}
     }
     \subfigure[][Sample of Comments]{
         \includegraphics[width=0.42\textwidth, height = 4.5cm]{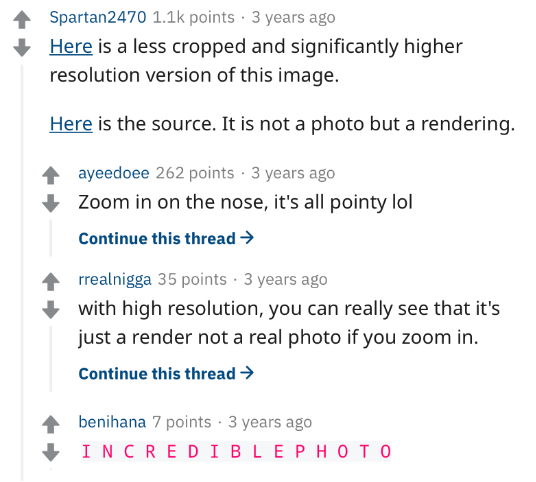}
        \label{fig:examplecomment}
     }
     \caption{\lanyu{Example of an Image-centric Post on Reddit}}
     \label{fig:example}
 \end{figure}

\begin{myDef}
    \emph{\textbf{Image-centric Post ($P$):} an image-centric post (Figure \ref{fig:example}) is a social media post that depicts an event, object, or topic with image(s), the context (i.e., text associated with the image), and  the comment section. } 
\end{myDef}

\begin{myDef}
    \emph{\textbf{Fauxtography (labeled as ``True"):} a post that conveys a misleading message to the viewers of the post. In particular, a post is a fauxtography if the image of the post i) directly supports a false claim, or ii) conveys misinformation of a true claim. }
\end{myDef}

\begin{myDef}
    \emph{\textbf{Non-Fauxtography (labeled as ``False"):} images that do not fall under ``fauxtography".}
\end{myDef}

To formulate our problem, we assume a set of $N$ posts $\mathcal{P} = \{P_1, P_2, ..., P_N\}$ from online social media. A post $P_n$, $1 \le n \le N$, is defined as a tuple: $P_n = (\mathcal{T}_n, \mathcal{I}_n, \mathcal{C}_n, y_n)$ where $\mathcal{T}_n$ and $\mathcal{I}_n$ refer to the text and the image part of the post, respectively.  $\mathcal{C}_n$ represents the comments (including shares and replies) of the post and  $y_n$ is the ground truth label on the fauxtography of $P_n$.


Given the above definitions, the goal of fauxtography detection is to classify each image-centric post into one of the two categories (i.e., fauxtography or not). Formally, for $P_n,  1 \le n \le N$, our goal is to find: 
\begin{equation} \label{eq:goal}
\operatorname*{arg\,max}_{\Tilde{y_n}} Pr(\Tilde{y_n} = y_n|P_n),~  \forall 1 \le n \le N
\end{equation} 
\noindent
where $\Tilde{y_n}$ denotes the estimated label for $P_n$.

\begin{figure}[!htb]
 \centering
     \subfigure[][Fauxtography]{
         \includegraphics[width=0.42\textwidth, height = 3cm]{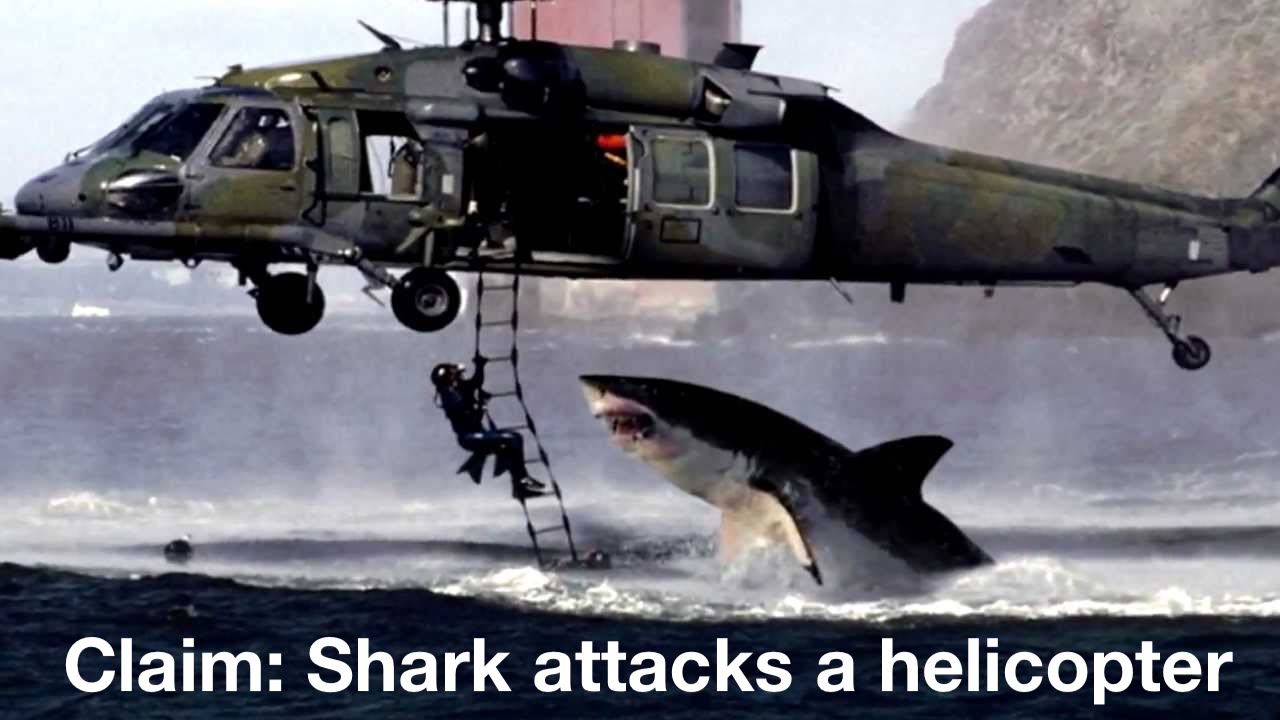}
         \label{fig:textf}
     }
     \subfigure[][Non-Fauxtography]{
         \includegraphics[width=0.42\textwidth, height = 3cm]{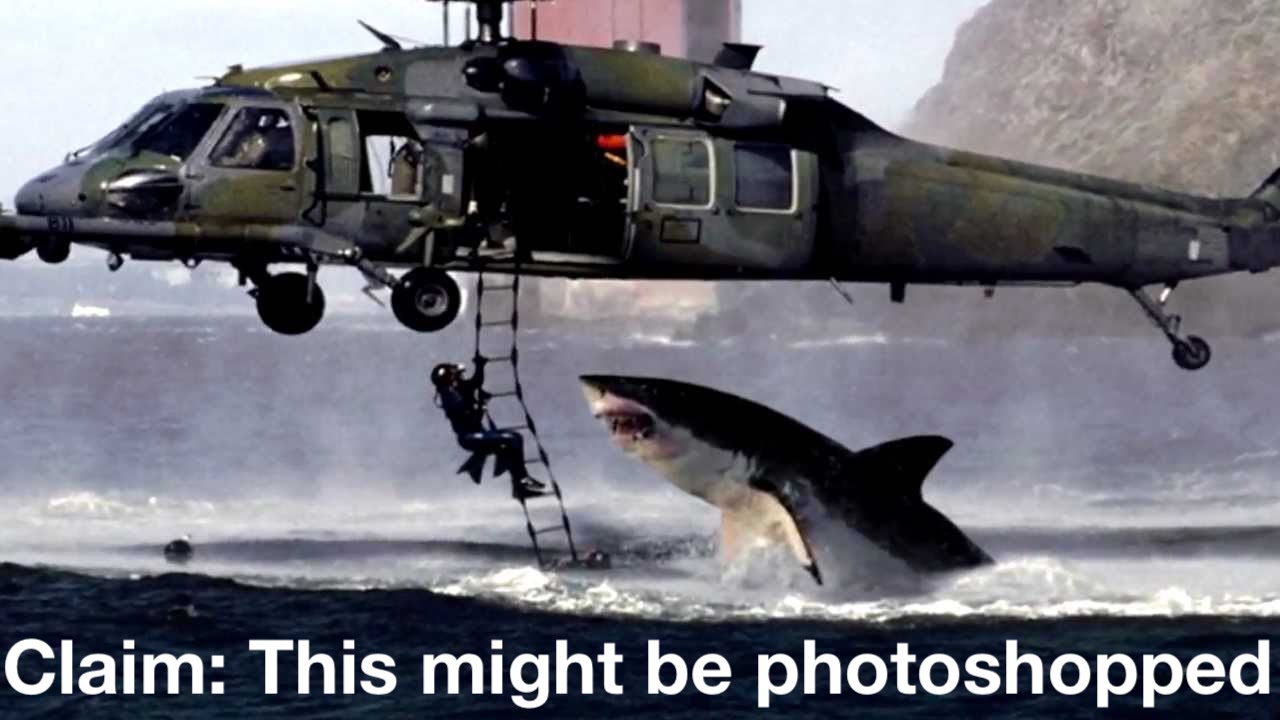}
        \label{fig:textt}
     }
     \caption{\lanyu{Example of Fauxtography and Non-fauxtography}}
     \label{fig:difftext}
 \end{figure}

Please note that the fauxtography detection problem is \emph{not equivalent to ``fake image"} detection \cite{gupta2013faking,huynh2016robust}, which only asserts whether the visual content of the image is manipulated or not. 
\lanyu{For example, Figure \ref{fig:difftext} shows two identical images.
The image itself is fake (i.e., it is created with photoshop), and should be classified as \textit{fake} by the ``fake image'' detection algorithm. However, in  the problem of fauxtography detection, posts with the same image could be classified into completely different categories when the image is accompanied by different claims as shown in \ref{fig:difftext}(a) and \ref{fig:difftext}(b).}
Also, fauxtography detection \emph{is not equivalent to ``false claim"} detection, which only focuses on checking the truthfulness of textual claims \cite{R1,vo2018rise}. The fauxtography detection requires a holistic analysis of the image and its associated context, which is a new research problem that has not been well addressed by current solutions.

\section{Solution}
\label{sec:solution}
 
\begin{figure}[!htb]
    \centering
    \includegraphics[width=0.96\textwidth]{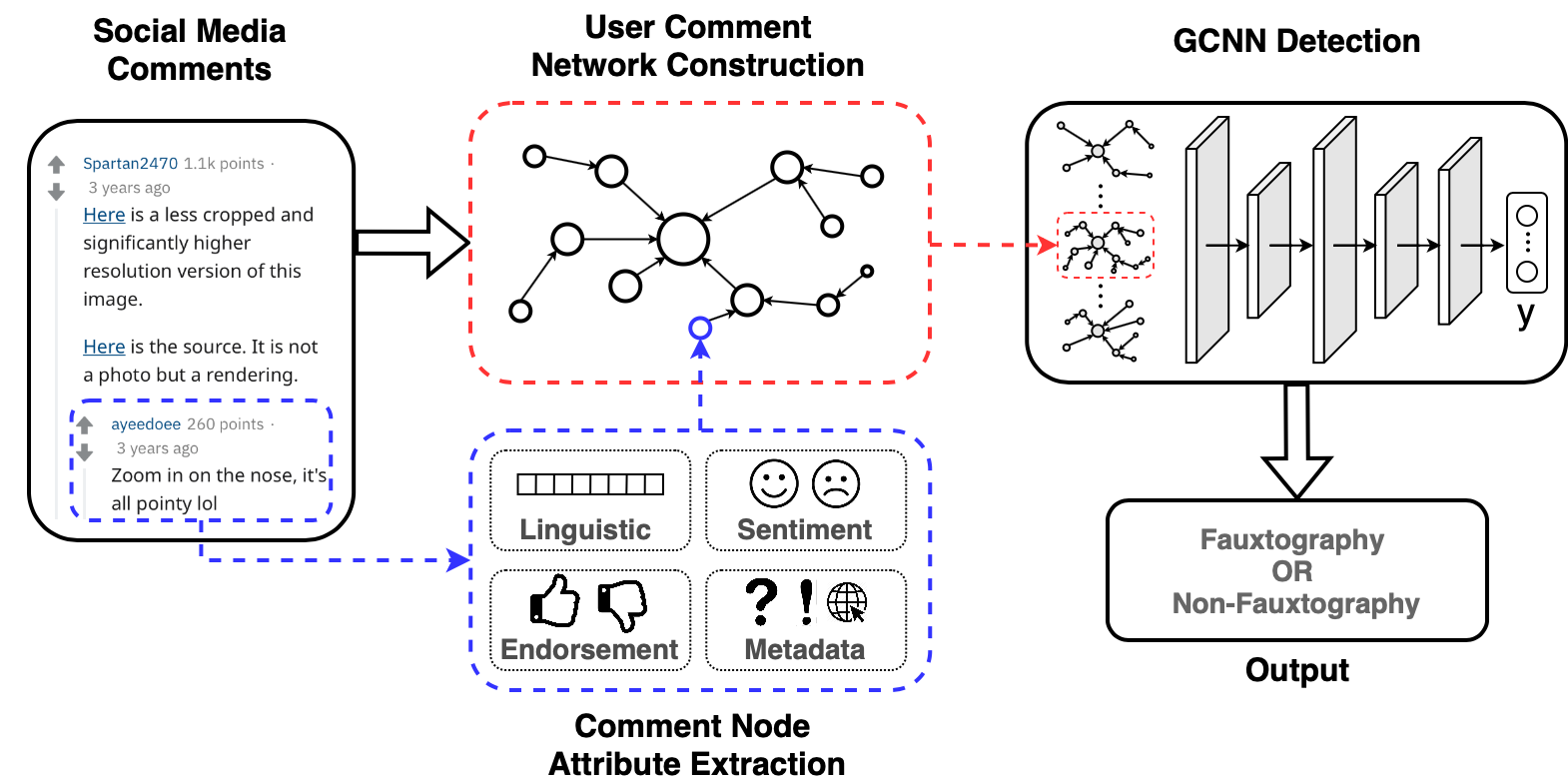}
    \caption{Overview of the FauxWard Framework}
    \label{fig:overview}
\end{figure}

In this section, we present the FauxWard framework to address the fauxtography detection problem formulated above. The FauxWard scheme is a graph convolutional neural network approach that leverages i) the topological characteristics underlying the user comment network of a social media post, and ii) the linguistic and semantic comment information extracted from the user comments.
An overview of the FauxWard framework is shown in Figure~\ref{fig:overview}. The FauxWard framework contains three major components: i) a \textit{User Comment Network Construction}
module that constructs the user comment network from the reply relationship of the comments associated with a post;
ii) a \textit{Comment Node Attribute Extraction}
module that extracts the complex information of the comment node with a set of linguistic and semantic attributes from each user comment;
iii) a \textit{GCNN Detection} module that jointly leverages the topological characteristics of the user comment network and the comment node information to classify the fauxtography posts through a principled graph convolutional neural network (GCNN) framework.
We will discuss the details below.

\lanyusrv{\subsection{User Comment Network Construction}\label{sec:sol:const}}

\shang{We first observe that fauxtography and non-fauxtography posts are different in terms of the topological structure of the user comment network (e.g., the length of a comment thread, the number of replies) and semantic features of user comments (e.g., emotion and polarity of user feedback).
For example, we found that users are more likely to use comments to show their negative attitude towards fauxtography posts (e.g., ``Aka, fake", ``wimpy"). These comments appear to be less attractive to other users for discussion, which often result in a large amount of single-comment threads. In contrast, non-fauxtography posts often get more engagement from social media users. 
To effectively capture the topological characteristics and semantic features of user comments, we model the comments of a social media post as a directed graph. Specifically, we first define a few key terms in our model.
}

\begin{myDef}
\label{def:network}
\shang{
    \emph{\textbf{User Comment Network $\mathbf{G}$:} the user comment network $\mathbf{G}$ of an image-centric post is constructed as a directed graph $\mathbf{G} = (\mathbf{V}, \mathbf{E})$, where $\mathbf{V}$ is a set of nodes, and $\mathbf{E}$ is a set of edges indicating the reply relationship between each pair of nodes.
    In particular, we define a \textit{source node} $v_0 \in \mathbf{V}$ to denote the content of the original social media post and other \textit{comment nodes} (i.e., $v_i \in \mathbf{V}, i \neq 0$) in the user comment network to represent the comments a post receives. We also define the \textit{edge} $e_{v_i,v_j}$ between two nodes $v_i$ and $v_j$ to denote the reply direction from comment $v_j$ to comment $v_i$.
    } 
    }
\end{myDef}

\begin{myDef}
\label{def:adjacency}
\lanyurv{
    \emph{\textbf{Adjacency Matrix $A$:} we also define an adjacency matrix $A \in \mathbb{R}^{V\times V}$ to record the edges between any pair of the comment nodes. Specifically, for all pair of nodes $v_i, v_j \in \mathbf{V}$, $A_{i,j} = 1$ if there is an edge $e_{v_i,v_j}$ between node $v_i$ and $v_j$, otherwise $A_{i,j} = 0$.} 
    }
\end{myDef}

\begin{figure}[!htb]
 \centering
     \subfigure[][Fauxtography]{
         \centering
         \includegraphics[width=0.42\textwidth]{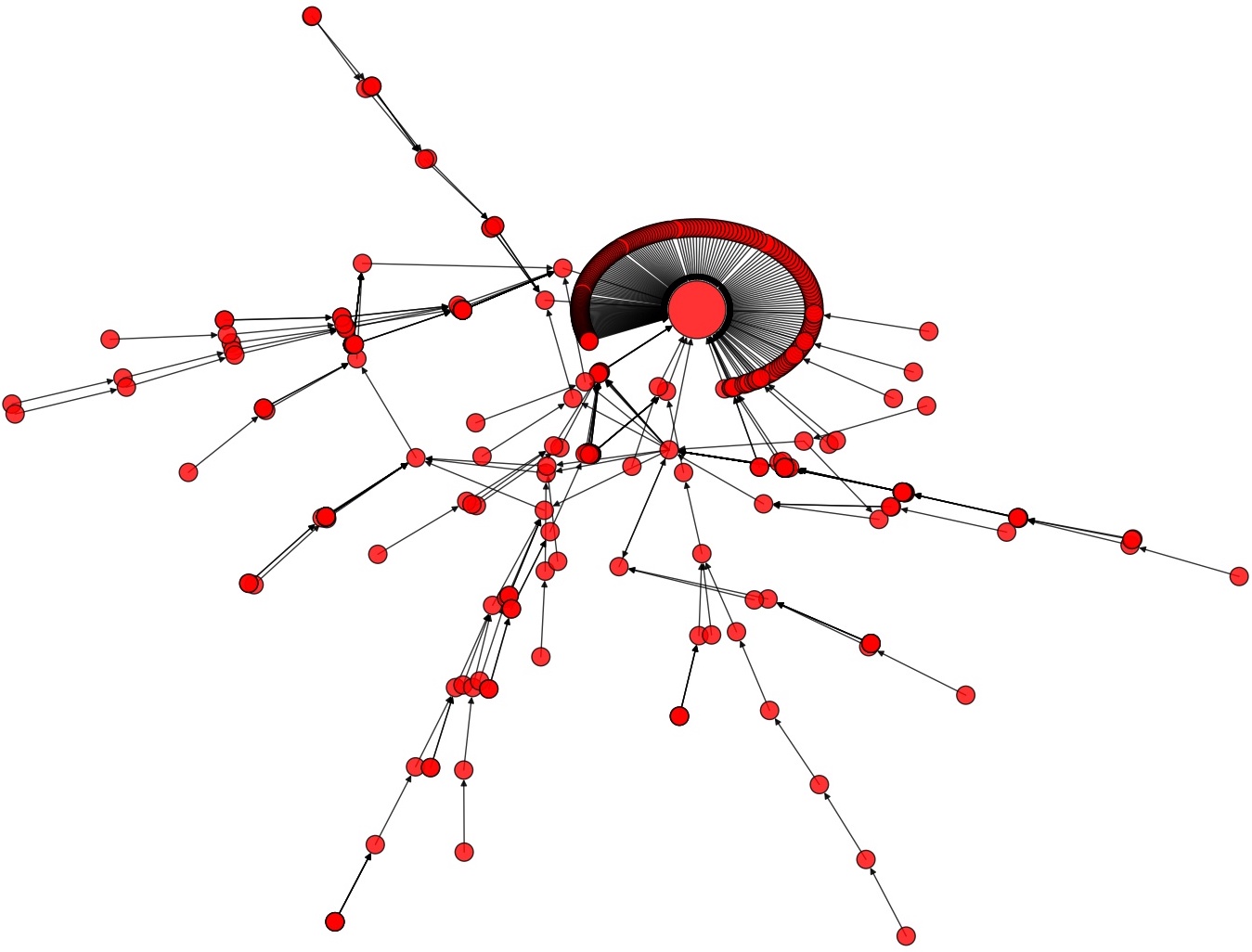}
         \label{fig:networkfaux}
     }
     \subfigure[][Non-Fauxtography]{
         \centering
         \includegraphics[width=0.42\textwidth ]{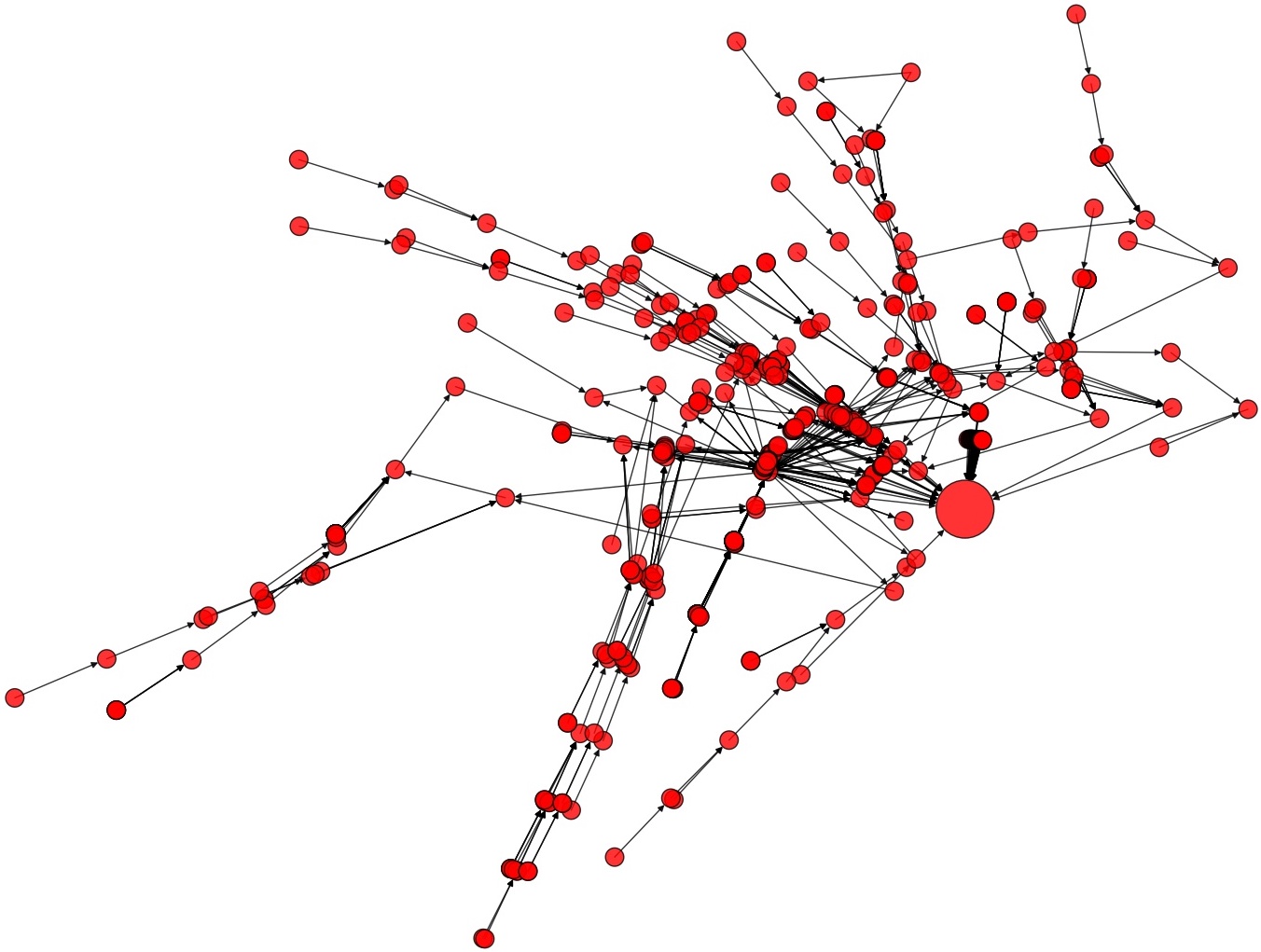}
        \label{fig:networknonfaux}
     }
     \caption{Examples of the User Comment Network for Fauxtography and Non-fauxtography Posts. The source node is denoted with large size.}
     \label{fig:network}
 \end{figure}

\lanyu{Figure \ref{fig:network} shows an example of the user comment network of a fauxtography and a non-fauxtography post. We observe that fauxtography posts often receive a large number of comments that directly reply to the post. In contrast, non-fauxtography posts often attract more subsequent discussion in the form of replies to a comment.}

\subsection{Comment Node Attribute Extraction}
\label{sec:sol:node}
\lanyusrv{We observe that the user comments often contain valuable information (e.g., the vocabulary used, the emotion and polarity reflected, and the endorsement or feedback from other users) in distinguishing fauxtography and non-fauxtography posts.} For example, the fauxtography post in Figure \ref{fig:fa} is likely to be debunked by a comment, ``Fake image! It is super easy to photoshop", and such a debunking comment is also likely to be appreciated and endorsed by other users in the form of like/dislike or retweets. 
\lanyusrv{Such a debunking comment can be captured by a comment node with negative polarity and high endorsement in the user comment network. Therefore, we extract a set of key features based on the empirical observation of user comments, and incorporate them into the structured user comment network constructed in Section \ref{sec:sol:const} to identify fauxtography posts.}
\shang{
In particular, we focus on a set of diversified comment node attributes (i.e., linguistic ($L_v$), sentiment~($S_v$), endorsement ($E_v$), and metadata ($M_v$)) in order to learn and represent the complex information embedded in each comment node $v$. 
We elaborate each comment node attribute below.}

\begin{myDef}
\lanyurv{
    \emph{\textbf{Linguistic Attribute $L_v$:} we define the linguistic attribute $L_v \in \mathbb{R}^{1\times K_L}$ of a comment node $v$ as a vector representation to represent the vocabulary used in each comment network.}
    }
\end{myDef}

\begin{figure}[!htb]
 \centering
     \subfigure[][Fauxtography]{
         \centering
         \includegraphics[width=0.42\textwidth, height = 3.2cm]{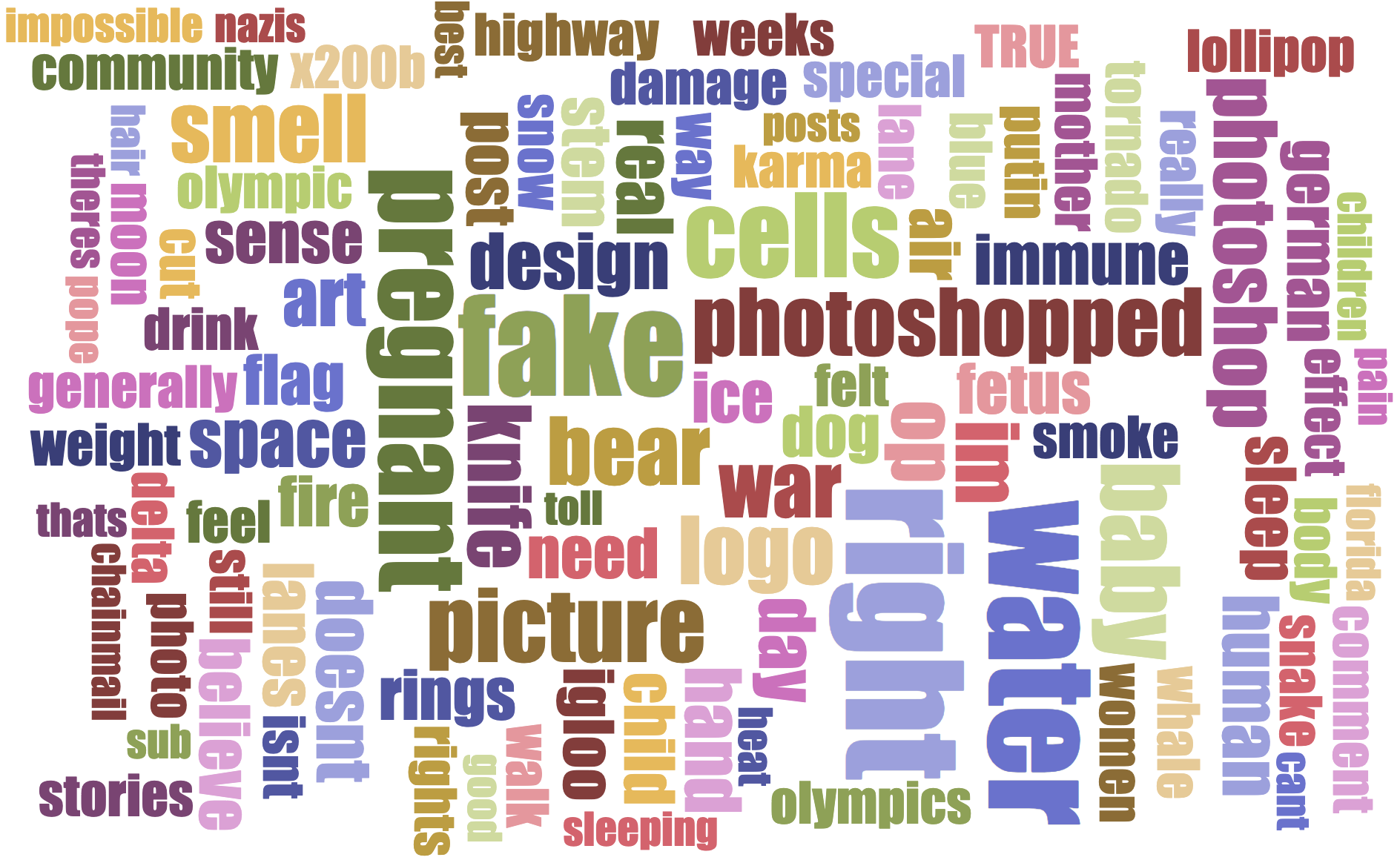}
         \label{fig:wcf}
     }
     \subfigure[][Non-Fauxtography]{
         \centering
         \includegraphics[width=0.42\textwidth, height = 3.2cm]{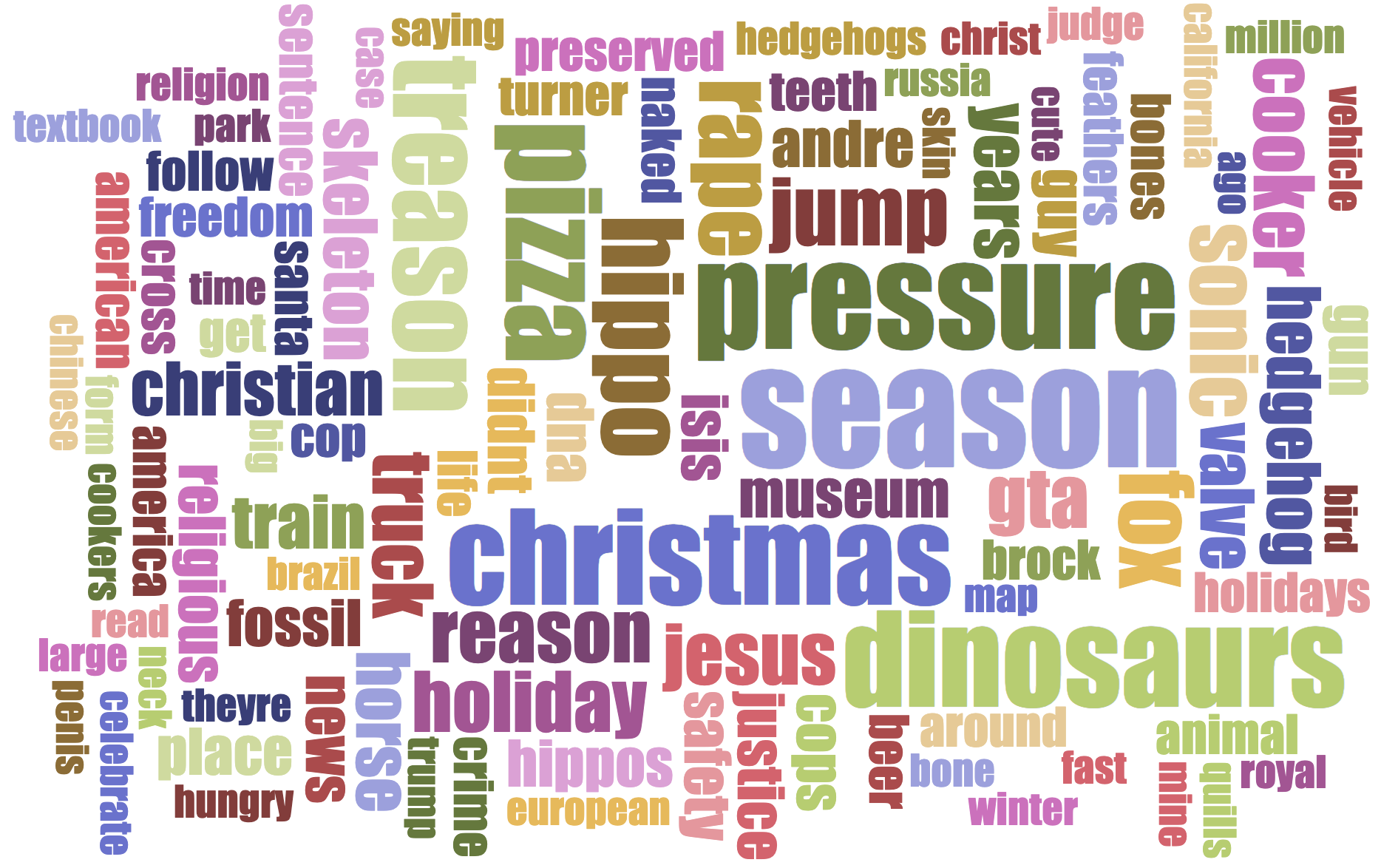}
        \label{fig:wct}
     }
     \caption{Word Cloud}
     \label{fig:wordcloud}
 \end{figure}

\lanyurv{An example of the the linguistic attribute is shown in Figure~\ref{fig:wordcloud}. We observe that vocabulary used in the comments of fauxtography and non-fauxtography posts are different to some extent.  In particular, (a) and (b) show the word clouds of comments in each post category. We note that image verity related words (e.g., ``photoshop", ``photo", ``fake'') appears more frequently in fauxtography posts. In contrast, comments in non-fauxtography posts contain more general news topics (e.g., ``treason", ``Christmas", ``ISIS"). }

\begin{myDef}
\lanyurv{
    \emph{\textbf{Sentiment Attribute $S_v$:} we define the sentiment attribute $S_v \in [-1.0, 1.0]$ of each comment node $v$ to be the polarity score to indicate the sentiment in each comment network. 
    Specifically, a positive polarity score (i.e., $S_v > 0$) indicates a positive sentiment, and vice versa.}
    }
\end{myDef}

\begin{figure}[!htb]
\centering
     \subfigure[][Fauxtography]{
         \centering
         \includegraphics[width=0.42\textwidth]{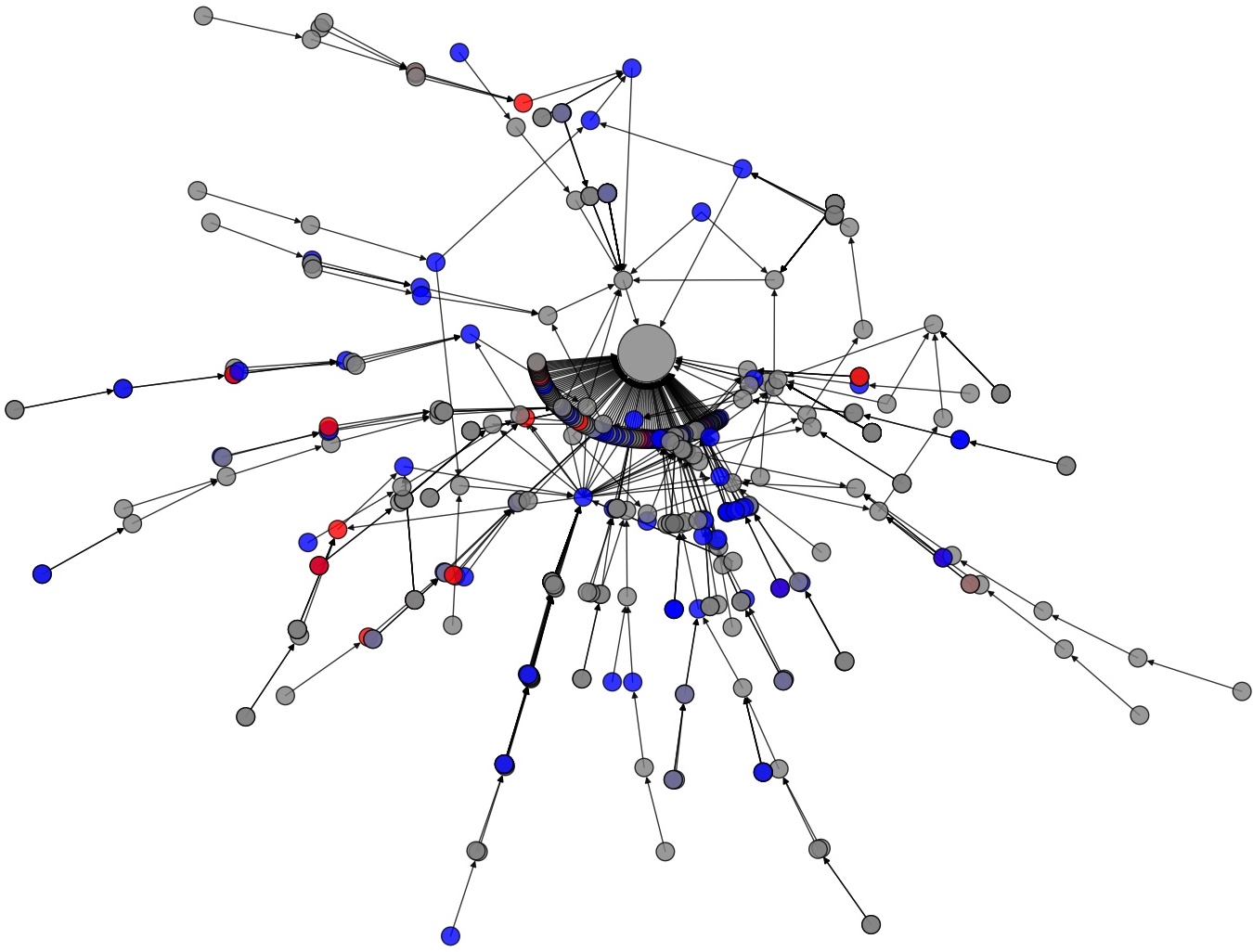}
         \label{fig:sentf}
     }
     \subfigure[][Non-Fauxtography]{
         \centering
         \includegraphics[width=0.42\textwidth]{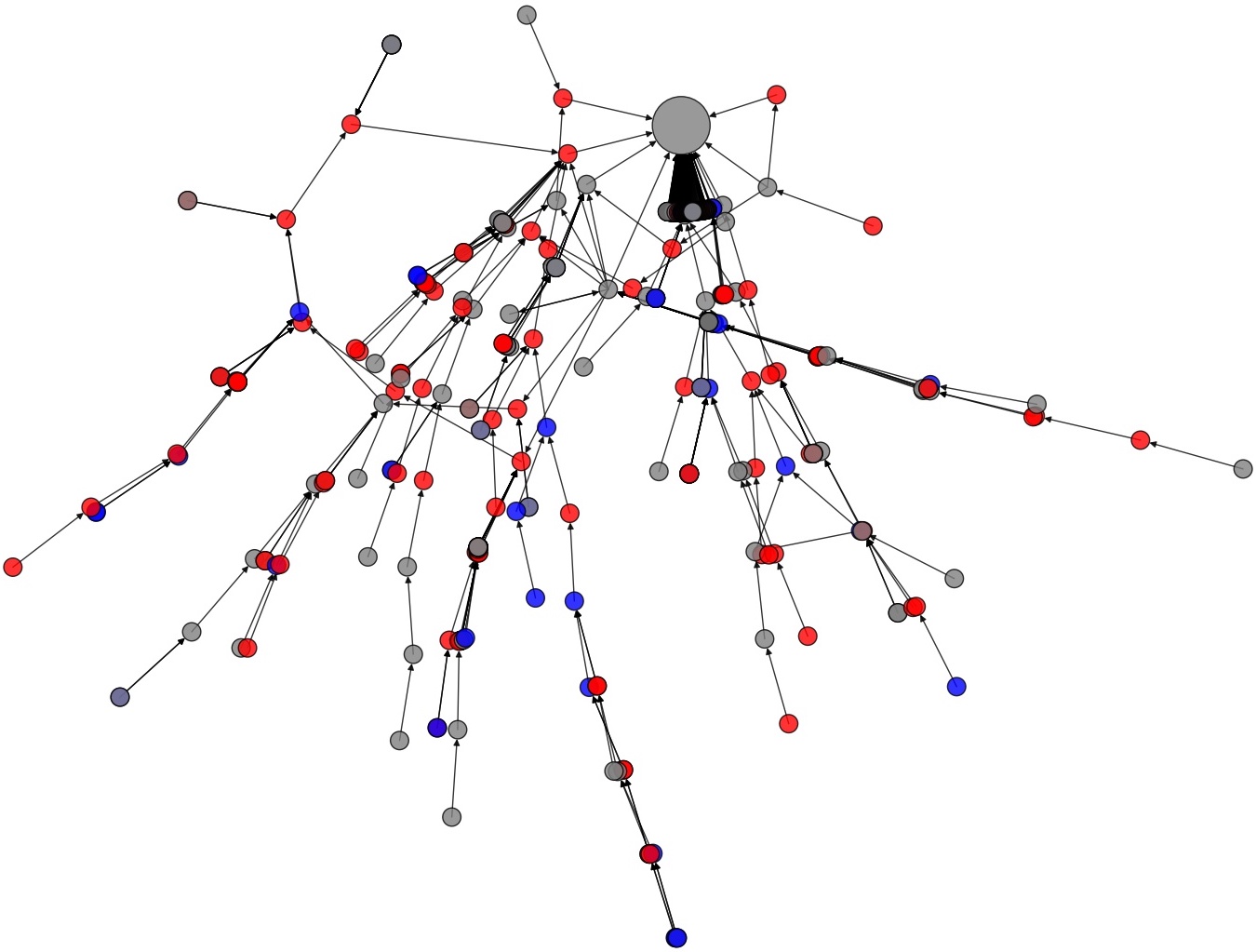}
        \label{fig:sentt}
     }
     \caption{\lanyu{Illustration of the Sentiment Attribute. The color of each comment node indicates the sentiment attribute of the corresponding comment, i.e., red - positive sentiment ($polarity\ge0.5$), blue - negative sentiment ($polarity\le-0.5$), grey - neutral sentiment ($0.5<polarity<0.5$).} }
     \label{fig:sentiment}
 \end{figure}

\lanyurv{Figure \ref{fig:sentiment} shows an example of the sentiment attribute in the user comment network of two social media posts. We observe that comments in the fauxtography posts often contain more negative ``echo chambers" (i.e., consecutive comments of negative sentiment) that indicate debunk and arguments between users, while the comment sentiments in non-fauxtography posts often appear to be more positive that reflect agreements from users. }

\begin{myDef}
\lanyurv{
    \emph{\textbf{Endorsement Attribute $E_v$:} we define the endorsement attribute $E_v \in \mathbb{R}$ as the number of aggregated endorsement a comment receives from other users. Specifically, $E_v$ equals to the number of likes - the number of dislikes for Reddit, and $E_v$ equals to the sum of the number of likes and the number of retweets for Twitter.}
    }
\end{myDef}

\begin{figure}[!htb]
    \centering
     \subfigure[][Fauxtography]{
         \centering
         \includegraphics[width=0.42\textwidth]{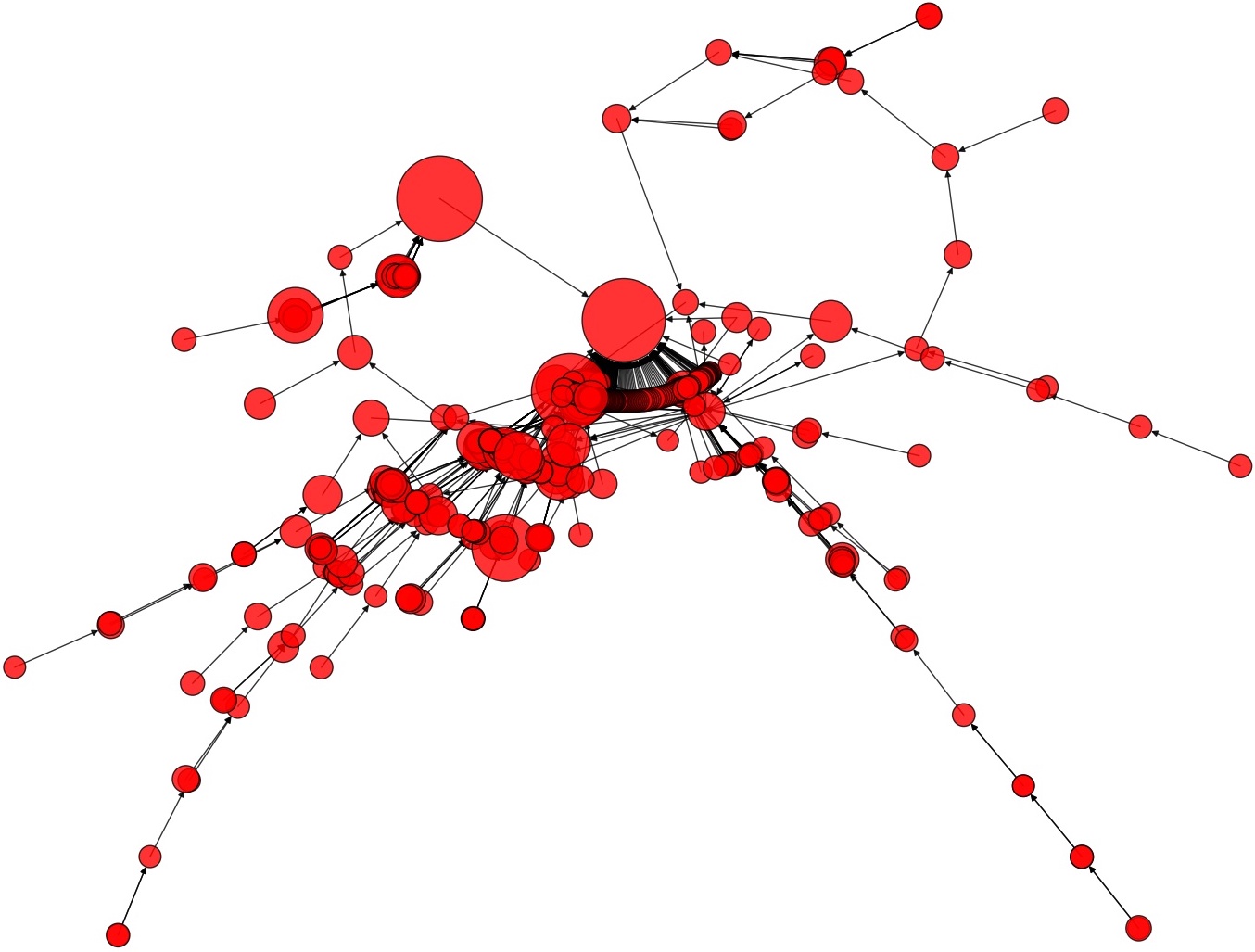}
         \label{fig:endorsef}
     }
     \subfigure[][Non-Fauxtography]{
         \centering
         \includegraphics[width=0.42\textwidth]{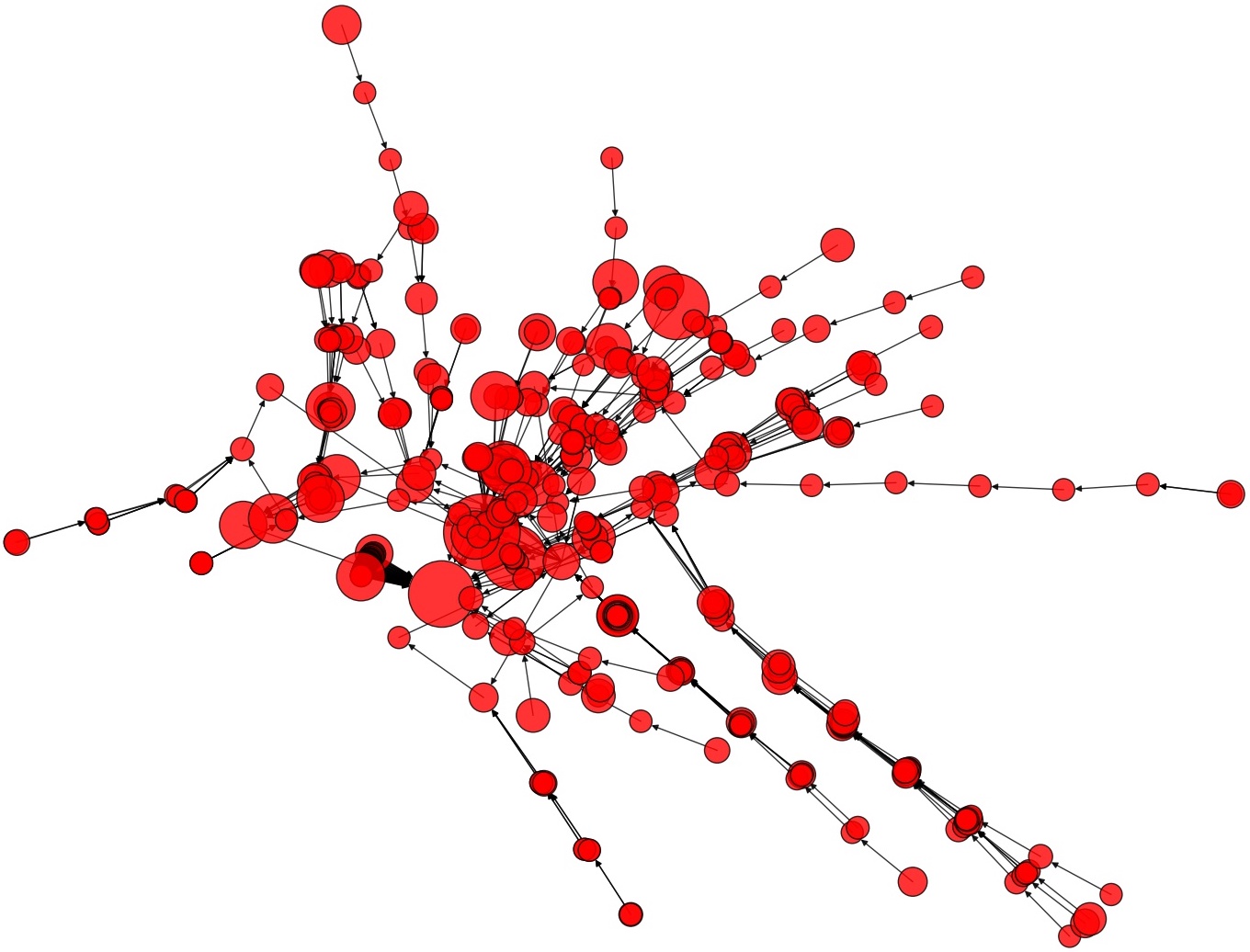}
        \label{fig:endorset}
     }
     \caption{\lanyu{Illustration of the Endorsement Attribute. The size of each comment node indicates the endorsement attribute (i.e., the number of aggregated likes) of the corresponding comment.}}
     \label{fig:endorsement}
 \end{figure}

\lanyurv{
Figure \ref{fig:endorsement} shows an example of the endorsement attribute in the user comment network. We observe that there are a few ``hub'' comments in the fauxtography post that receives a large amount of support (i.e., endorsement) from other users. Such ``hub'' comments are often the ones that directly debunk the fauxtography in the post and thus receives support from users sharing similar points of view. In contrast, the endorsement attribute of comments in non-fauxtography posts appears to be more diversified as users often pay more attention to the content beyond the truthfulness of the image in those scenarios. }

\begin{myDef}
\lanyurv{
    \emph{\textbf{Metadata Attribute $M_v$:} we define the metadata attribute $M_v \in \mathbb{R}^{1 \times K_M}$ as a set of metadata features extracted from each comment node $v$.}
    }
\end{myDef}

\lanyurv{These metadata features are often shown direct correlations to characterizing the fauxtography posts.
For example, we observe that fauxtography posts often debunked by comments containing URLs that link to the original image or the true story associated with the image. We also observe that comments of fauxtography posts often contain many verity-related (e.g., ``fake", ``false alarm") or image-related words (e.g., ``photo", ``photoshop"). A summary of the extracted metadata features is listed in Table \ref{tab:metafeatures}.  }

\begin{table}[htb!]
    \centering
    \small
    \caption{Metadata Attribute}
    \label{tab:metafeatures}
    \centering
    \begin{tabular}{l l }
        \toprule
        \midrule
        \textbf{Feature} &  \textbf{Description}  \\
        \hline
        \texttt{Word Count} & Number of words in a comment \\
        \hline
        \texttt{Verity Terms} & Number of verity-related terms in a comment \\
        \hline
        \texttt{Image Terms} & Number of image-related terms in a comment  \\
        \hline
        \texttt{Question Marks} & Number of question marks in a comment\\
        \hline
        \texttt{Exclamation Marks} & Number of exclamation mark in a comment \\
        \hline
        \texttt{URLs} & Number of URLs in a comment \\
        \midrule
        \toprule
    \end{tabular}
\end{table}

\shang{Finally, we define the node feature vector to represent the comment node attributes that contain the key characteristics of each user comment. }

\begin{myDef}
\label{def:node}
\lanyurv{
    \emph{\textbf{Node Feature Vector $F_v$:} 
    \shang{the node feature vector $F_v$ for a comment node $v$ is defined as $F_v = [L_v, S_v, E_v, M_v], \text{s.t. } F_v \in \mathbb{R}^{1\times K} ~\forall ~ v \in \mathbf{V}$ and $K$ is the sum of the dimensions of node attributes.} 
    We denote the node feature matrix $F \in \mathbb{R}^{V\times K}$ as the matrix that stores the feature vectors for all nodes in the user comment networks. 
    }}
\end{myDef}

\subsection{GCNN Detection}

\dz{
In FauxWard, we model the fauxtography detection task as a graph classification problem and develop a novel graph convolutional neural network approach to solve it.}
\yz{A key challenge of our graph convolutional neural network design lies in effectively characterizing and encoding the user comment networks defined in Section \ref{sec:sol:const} and \ref{sec:sol:node}. 
\lanyusrv{On one hand, the user comment networks (Definition \ref{def:network}) are different in terms of their topological features (e.g., different sizes and structures of the user comment networks).
On the other hand, each comment node (Definition \ref{def:node}) consists of diversified node attributes with distinct linguistic and semantic representations.}}
\dz{In light of such a challenge, we design a graph convolution neural network model that jointly leverages i) the topological characteristic of the user comment networks, and ii) the rich linguistic and semantic attributes of the comments to detect fauxtography posts.}
\lanyurv{To address the topological challenge in the user comment networks, we design a cluster-based pooling layer in the GCNN framework that first clusters neighboring nodes of various sizes based on their node embeddings, and \lanyusrv{reconstructs the input graphs to the next graph convolutional layer in our model.}
\lanyusrv{In addition, we take advantage of the vector representation of each comment node and encode the diversified comment node attributes into the user comment network to classify fauxtography posts.}}
We present the details of our approach below.

\begin{figure}[!htb]
    \centering
    \includegraphics[width=0.98\textwidth]{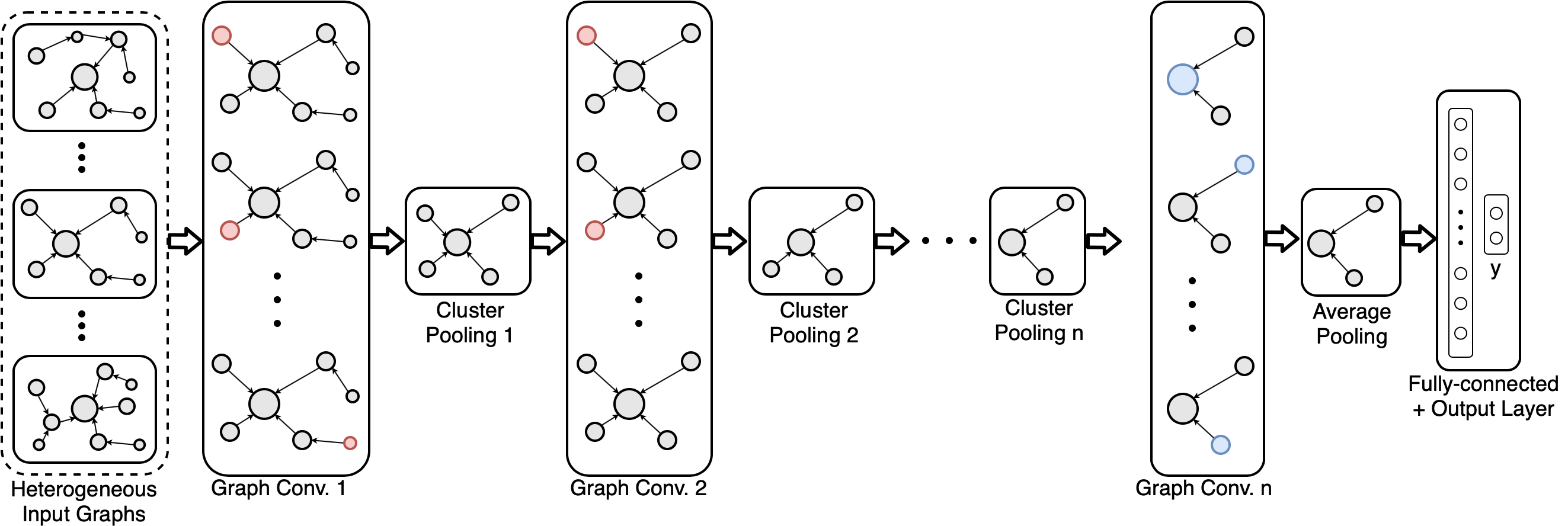}
    \caption{The architecture of the GCNN Detection Module}
    \label{fig:gcn}
\end{figure}

An overview of the architecture of the GCNN detection framework is summarized in Figure \ref{fig:gcn}. 
Let $\mathcal{G = \{\mathbf{G_1}, \mathbf{G_2}, \cdots, \mathbf{G_N}\}}$ be a collection of image-centric social media posts, where 
$\mathbf{G_n}$, $1 \leq n \leq N$, is the user comment network of post $P_n$ (as defined in Section \ref{sec:sol:const}). $y_n$ and $\Tilde{y_n}$ are the corresponding ground truth and estimated labels of the post, respectively. For a given set of $N$ social media posts $\mathcal{P} = \{(\mathbf{G_1}, y_1), (\mathbf{G_2}, y_2), \cdots, (\mathbf{G_N}, y_N)\}$, we aim to find: 
 \begin{equation}
    \operatorname*{arg\,max}_{\Tilde{y_n}} Pr(\Tilde{y_n} = y_n|\mathbf{G_n}),~  \forall ~ 1 \le n \le N
 \end{equation}

\shang{A key challenge in our problem is to effectively extract topological features of user comment network with various sizes (i.e., different number of nodes and edges). To address the challenge, we first define a key concept as follows.
\begin{myDef}
\emph{\textbf{Unified User Comment Node Space $\mathcal{V}$:} the union of all comment nodes in the collection of posts. Formally, $\mathcal{V} = \bigcup_{n=1}^{N} \mathbf{V_n} ~ \forall ~ 1 \leq n \leq N$, where $\mathbf{V_n}$ is the set of nodes in user comment network $\mathbf{G_n}$. The size of $\mathcal{V}$ is denoted as $V$. An edge between any pair of comment nodes in $\mathcal{V}$ can be recorded using the corresponding adjacency matrix as defined in Definition \ref{def:adjacency}. } 
\end{myDef}
Then, any user comment network $\mathbf{G} \in \mathcal{G}$ can be represented as $(A, F)$ where $A \in \mathbb{R}^{V\times V}$ is the adjacency matrix and $F \in \mathbb{R}^{V\times K}$ is the node feature matrix with respect to the unified user comment node space $\mathcal{V}$. However, a direct representation of the user comment network $\mathbf{G}$ with respect to $\mathcal{V}$ will result in a large and sparse adjacency matrix $A$. Therefore, we adopt the ``block diagonal adjacency matrix'' strategy to effectively handle the sparse and various sized user comment networks. Formally, let $\bar A_1, \bar A_2, \cdots, \bar A_M$ be the non-empty adjacency matrices (i.e., adjacency matrices without empty rows and columns) for input graphs $\mathbf{G_1}, \mathbf{G_2}, \cdots, \mathbf{G_M}$, the block diagonal adjacency matrix $A_{diag}$ for $M$ input graphs is defined as: 
\begin{equation}
    A_{diag} = 
    \begin{bmatrix}
    \bar A_1 & & &\\
    &   \bar A_2 & & \\
    & & \ddots &\\
    & & & \bar A_M \\
    \end{bmatrix}
\end{equation}
We then perform sparse matrix multiplication with respect to the $A_{diag}$ to efficiently train the model with batch-wise training. 
}

Next, we adopt the recursive neighborhood aggregation (or ``message-passing") strategy in the graph convolutional layer as follows:
\lanyurv{
\begin{equation}
    H^{(k)} = f(A^{(k-1)}, H^{(k-1)}, W^{k-1})
\end{equation}
where $A^{(k-1)}$ and $H^{(k-1)}$ are the input node adjacency matrix and node feature matrix at the $k^{th}$ layer of the GNN, respectively. $W^{k}$ is the trainable weighting parameters and $f(\cdot)$ is the message propagation function. In particular, we initialize the graph convolutional neural network with the user comment networks we constructed in Section \ref{sec:sol:const} (i.e., $A^{(0)}=A$), and the set of comment node attributes we extracted in Section \ref{sec:sol:node} (i.e., $H^{(0)}=F$). Formally, 
\begin{equation}
    H^{(1)} = f(A, F, W^{0})
\end{equation}
}

To aggregate node information in the GCNN framework, we apply graph convolutional layer to the neighbor nodes $\mathcal{N}(v)$ of each node $v$ in the graph, and use the rectified linear unit (ReLU) as the activation function $\sigma(\cdot)$. 
However, the user comment networks for social media posts often appear to be large and various sized in terms of the number of nodes and edges, which will result in a large and sparse adjacency matrix and cause the potential gradient vanishing problem. To this end, we applied the first-order approximation of localized spectral filters on graph \shang{convolutional} layer~\cite{kipf2016semi} with added self-connection of the adjacency matrix. 
Formally, the updated adjacency matrix in each graph convolutional layer is formulated as: 
\begin{equation}
\Tilde{A}^{(k)} = \hat{D}^{(k)^{-\frac{1}{2}}}\hat{A}^{(k)}\hat{D}^{(k)^{-\frac{1}{2}}}
\end{equation}
where $\hat{A}^{(k)} = I+A^{(k)}$ is the adjacency matrix with added self-loops and $I$ is the identity matrix. $\hat{D}$ is the diagonal degree matrix where $\hat{D}_{ii} = \sum_{j}\hat{A}_{ij}$.
Formally, the $k^{th}$ graph convolutional layer is defined as:
\begin{equation}
\begin{split}
    H^{(k)} &= f\left(\Tilde{A}^{(k-1)}, H^{(k-1)}, W^{k-1}\right)\\
            &= \sigma \left(\hat{D}^{(k-1)^{-\frac{1}{2}}}\hat{A}^{(k-1)}\hat{D}^{(k-1)^{-\frac{1}{2}}}H^{(k-1)}W^{(k-1)} \right)
\end{split}
\end{equation}

\lanyu{In addition, we add a cluster-based pooling layer 
between the graph convolutional layers to coarsen the graph and efficiently learn the graph representation through the GCNN framework~\cite{ying2018hierarchical}. The cluster-based pooling layer first assigns neighboring nodes into clusters according to node embeddings learned from the previous graph convolutional layer and learns a representation for each cluster that is the input of the next graph convolutional layer.} 
\lanyurv{Let $C^{(k)}$ be the clustering matrix after the $k^{th}$ graph convolutional layer. We update the adjacency matrix $\Tilde{A}^{(k)}$ and node feature matrix $H^{(k)}$ as follows:
\begin{equation}
\Tilde{A}^{(k)} = C^{(k-1)^T} \Tilde{A}^{(k-1)} C^{(k-1)}
\end{equation}
\begin{equation}
H^{(k)} = C^{(k-1)^T} f\left(\Tilde{A}^{(k-1)}, H^{(k-1)}, W^{k-1}\right)\\
\end{equation}
}

\lanyurv{In this way, we can efficiently extract and preserve the topological features of local substructure (i.e., clusters) in the user comment network. Moreover, such a clustering design of the GCNN can also help to effectively extract and aggregate node information in the user comment network and high-level graph representations, especially for the posts with a large number of comments~\cite{ying2018hierarchical}.}

\lanyu{Finally, we use mean pooling as the readout layer to summarize the hidden graph representation before the fully-connected layer. A softmax layer is the last layer to output the binary classification results. We adopt the Adaptive  Moment Estimation (Adam) optimizer \cite{kingma2014adam} to train the graph neural network and minimize the cross-entropy loss:
\begin{equation}
   \mathcal{L} = - \frac{1}{N}\sum_{n=1}^{N}\left(y_n \log \Tilde{y_n} + (1-y_n) \log(1-\Tilde{y_n}) \right)
\end{equation}
}


\section{Data}
\label{sec:data}
\lanyu{
In this section, we describe the real-world dataset collected from the leading online social media platform Reddit\footnote{https://www.reddit.com/} and Twitter \footnote{https://www.twitter.com/}.  
Reddit, self-described as ``front page of the Internet", is a popular news aggregation site \cite{priya2019should} where massive fresh internet content is constantly shared and commented on by its users.
As of October 2019, Reddit has 430 million monthly active users, 199 million posts, and 1.7 billion comments~\cite{redditstats}. 
Twitter is a global micro-blogging platform hosting 330 million active users and 500 million visitors each month~\cite{twitterstats}.
}

We observe that both Reddit and Twitter have a huge amount of posts that are image-based. It is challenging to collect ground-truth labels for fauxtography posts on these media platforms. To address such a challenge, we first collect verified fauxtography images from 3 independent fact-checkers (i.e., snopes.com, factcheck.org, truthorfiction.com) in a similar way as \cite{lazer2018science}. The ground truth labels are initially decided based on the majority vote of these fact-checkers. We then assign three independent annotators to manually verify the label of each post using databases of historical facts and Google search. \lanyu{The dates of the fack-checked fauxtography images range from January 2010 to October 2019.} 

Given the labeled images, we perform a reverse search using the Google Vision API \footnote{https://cloud.google.com/vision/} to identify the original web URLs that contain the image. If the URL points to a social media post on Reddit or Twitter, we crawl the post and its comment threads using a crawler script we developed. We summarize the real-world dataset used for evaluation in Table \ref{table:datatrace}. 
We observe that 
there is a non-trivial amount of the fauxtography posts (10.6\% in Reddit and 11.3\% in Twitter) actually contain real images. This observation validates the unique challenge of fauxtography detection, where real images can also be leveraged to convey misleading messages.

\begin{table}[htb!]
    \small
    \caption{Data Trace Statistics}
    \label{table:datatrace}
    \centering
    \begin{tabular}{l c c}
        \toprule
        \midrule
        \textbf{Data Trace} &   \textbf{Reddit}  & \textbf{Twitter} \\
        \hline
        Number of Fact-checked Posts & 220 & 438\\
        \hline
        Number of Fauxtography &  179  & 378\\
        \hline
        Number of Fauxtography with Real Images &19  &  43\\
        \hline
        Number of Comments & 64,183 & 1,125,622 \\
        \hline
        Number of Distinct Users & 40,806 & 447,897\\
        \midrule
        \toprule
    \end{tabular}
\end{table}

We observe that the social media posts collected from the fact-checking websites are often biased (e.g., there are more fauxtography posts than non-fauxtography ones).  \dz{To mitigate this issue, we design a new data collection strategy. In particular,  for each fact-checked post found on Reddit, we collect 20 posts immediately ahead and behind the post in the same subreddit (i.e., sub-forum under the same topic) on the same day so that the collected posts reflect the actual ratio of fauxtography on that subreddit}.  Similarly, for each fact-checked post found on Twitter, we randomly sample 20 tweets that published within the same day as the fact-checked post. Removing invalid posts that do not contain image content, we finally obtain the datasets of 2780 and 2875 posts for Reddit and Twitter respectively, and assume all the posts are not fauxtography.

\begin{table}[htb!]
    \small
    \caption{\lanyu{Supplementary Data Trace Statistics}}
    \label{table:supdatatrace}
    \centering
    \begin{tabular}{l c c}
        \toprule
        \midrule
        \textbf{Data Trace} &   \textbf{Reddit} &\textbf{Twitter}  \\
        \hline
        Number of Posts &   2,780 & 2,875\\
        \hline
        Number of Comments & 395,964 & 2,205,635\\
        \hline
        Number of Distinct Users & 141,034 & 912,956 \\
        \midrule
        \toprule
    \end{tabular}

\end{table}


\section{Evaluation} \label{sec:eval}
In this section, we evaluate the FauxWard scheme using the real-world online social media datasets described in the previous section. \lanyu{We compare the detection performance of FauxWard with state-of-the-art baseline solutions as well as the FauxBuster solution in our previous work. The results show that the FauxWard scheme significantly outperforms all compared baselines in terms of detection accuracy and efficiency.}

\subsection{Baselines}
We compare the FauxWard with state-of-the-art baselines in fake image detection and fake claim detection.
\begin{itemize}
   
    \item  \lanyu{\textit{FauxBuster:} A random walk based network embedding solution particularly designed to detect fauxtography posts on social media using user comments~\cite{zhang2018fauxbuster}.}
    
    \item \textit{Fake Image:} A feature engineering based approach to detect fake images on social media using a decision tree classifier \cite{gupta2013faking}.
    
    \item \shang{\textit{SAME:} A deep learning based framework to detect multimodal fake news by leveraging features extracted from the news content and the sentiment of user comments~\cite{cui2019same}.} 
    
    \item \textit{Truth Discovery:}  A representative fact-checking scheme to detect misinformation among conflicting text-based claims on social media~\cite{zhang2018scalable}.
    
    \item \lanyu{\textit{Fake News:} A linguistic-based approach to identify fake news by extracting lexical and syntactic features from the news statement~\cite{shu2017fake}.}

    \end{itemize}
\lanyu{Please note that we carefully tune parameters in each baseline model to achieve its optimal performance for a fair comparison with the proposed scheme. In particular, for all of the compared methods, we use 80\% of the evaluation dataset as the training set and tune parameters based on the 5-fold cross-validation performance on the training set. }

\subsection{Detection Effectiveness}
\lanyu{
In the first set of experiments, we evaluate the detection effectiveness of FauxWard and the aforementioned baseline solutions. In particular, we adopt the commonly used metrics for binary classification evaluation, including \textit{Accuracy}, \textit{Precision}, \textit{Recall}, and \textit{F1-score}.}
\lanyu{The results are summarized in Table \ref{tab:classification_reddit} and Table \ref{tab:classification_twitter}. We observe that FauxWard significantly outperforms all the baseline schemes.
 \shang{In particular, on the Reddit dataset, FauxWard achieves a performance gain of 15.1\%, 9.1\%, 14.1\%, 18.4\%, and 32.5\% in terms of F1 score compared to the \textit{FauxBuster}, \textit{Fake Image}, \textit{SAME}, \textit{Truth Discovery}, and \textit{Fake News} baselines, respectively. On the Twitter dataset, FauxWard outperforms the \textit{FauxBuster}, \textit{Fake Image}, \textit{SAME}, \textit{Truth Discovery}, and \textit{Fake News} baselines by 6.7\%, 18.9\%, 9.39\%, 19.3\%, and 36.9\% in terms of F1 score, respectively.}}

\lanyu{
We observe that our FauxWard scheme did outperform the previous FauxBuster scheme. This is because FauxBuster takes users' comments as a whole document to extract the linguistic features (i.e., document embedding), which under-explores the topological patterns underlying the user comment network during the representation learning process. In contrast, FauxWard aggregates the linguistic attribute as well as semantic attributes of each comment through a GCNN framework to preserve such topological patterns.
Moreover, the Fake Image baseline also fails to detect fauxtography posts effectively because it only focuses on image features but does not put them into the context of the textual claims. Therefore, it is not robust against the fauxtography posts containing real images. In addition, the Truth Discovery and Fake News schemes only consider whether the textual claims are truthful or not. This leads to nontrivial false negatives in the results (i.e., fauxtography with fake images but truthful textual claims). In contrast, FauxWard is explicitly developed to detect the fauxtography posts by considering both the image and textual claim together with the message that they collectively express.} The results again demonstrate that existing image forgery detectors and fact-checkers cannot effectively solve the fauxtography detection problem.

\begin{table}[htb!]
    \small
    \centering
    \caption{Classification Accuracy for All Schemes (Reddit)}
    \begin{tabular}{l c c c c }
        \toprule
        \midrule
        \textbf{Algorithm}&\textbf{Accuracy}&\textbf{Precision}&\textbf{Recall}&\textbf{F1-Score}\\
        \cmidrule(l){1-5}
        \textbf{FauxWard} &\textbf{0.7536}& \textbf{0.7895}& \textbf{0.7692}& \textbf{0.7793} \\
        \cmidrule(l){1-5}
        \textbf{FauxBuster} &0.6812& 0.6216& 0.7419& 0.6765  \\
        \cmidrule(l){1-5}
        \textbf{Fake Image} &0.6522& 0.6667& 0.7692& 0.7143 \\
        \cmidrule(l){1-5}
        \shang{\textbf{SAME}} &\shang{0.6232} & \shang{0.6364} & \shang{0.7368} & \shang{0.6829} \\
        \cmidrule(l){1-5}
        \textbf{Truth Discovery} &0.6087& 0.6047& 0.7222& 0.6582\\
        \cmidrule(l){1-5}
        \textbf{Fake News} &0.5942 & 0.6061 & 0.5714 & 0.5882 \\
        \midrule
        \toprule
    \end{tabular}
\label{tab:classification_reddit}
\end{table}

\begin{table}[htb!]
    \small
    \centering
    \caption{Classification Accuracy for All Schemes (Twitter)}
    \begin{tabular}{l c c c c }
        \toprule
        \midrule
        \textbf{Algorithm}&\textbf{Accuracy}&\textbf{Precision}&\textbf{Recall}&\textbf{F1-Score}\\
        \cmidrule(l){1-5}
        \textbf{FauxWard} &\textbf{0.7109}& \textbf{0.7015}& \textbf{0.7344}& \textbf{0.7176} \\
        \cmidrule(l){1-5}
        \textbf{FauxBuster} &0.6797 & 0.6885 & 0.6562 & 0.6720  \\
        \cmidrule(l){1-5}
        \textbf{Fake Image} &0.6094 & 0.6129 & 0.5938 & 0.6032 \\
        \cmidrule(l){1-5}
        \shang{\textbf{SAME}} &\shang{0.6641} & \shang{0.6721} & \shang{0.6406} & \shang{0.6560} \\
        \cmidrule(l){1-5}
        \textbf{Truth Discovery} &0.6042 & 0.6056 & 0.5972 & 0.6014\\
        \cmidrule(l){1-5}
        \textbf{Fake News} &0.5312 & 0.5323 & 0.5156 & 0.5238 \\
        \midrule
        \toprule
    \end{tabular}
\label{tab:classification_twitter}
\end{table}

\begin{figure}[!htb]
    \centering
    \includegraphics[width=0.8\linewidth]{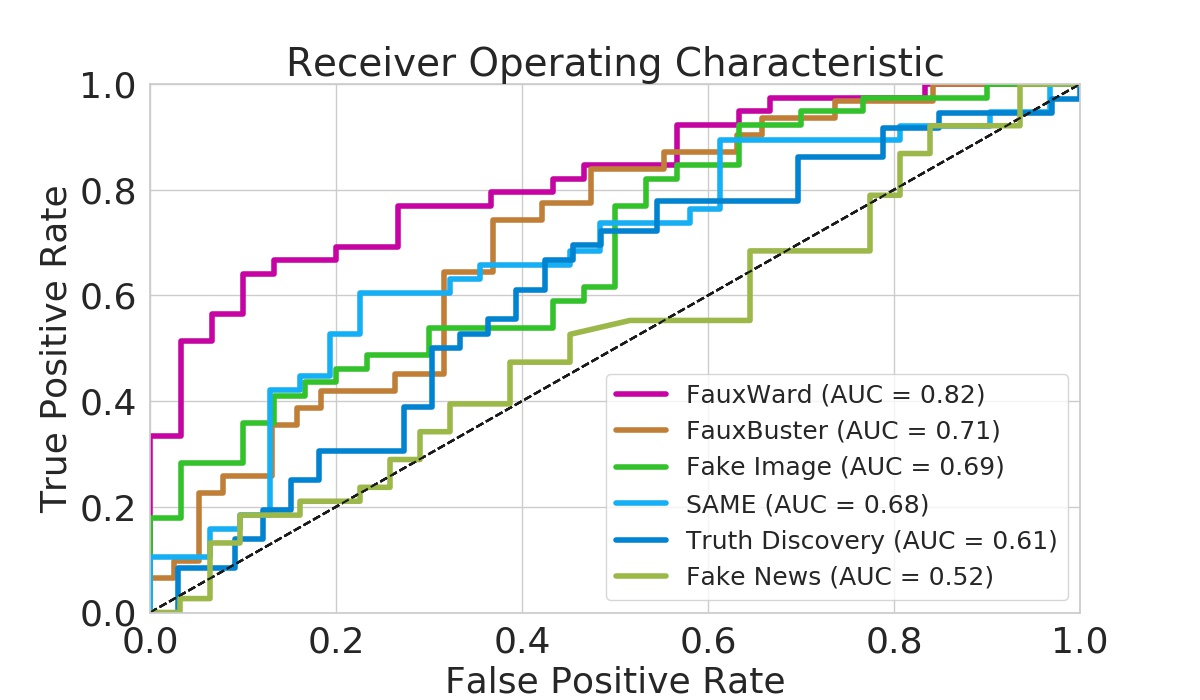}
    \caption{\shang{ROC Curve of All Schemes (Reddit)}}
    \label{fig:rocr}
\end{figure}

\begin{figure}[!htb]
    \centering
    \includegraphics[width=0.8\linewidth]{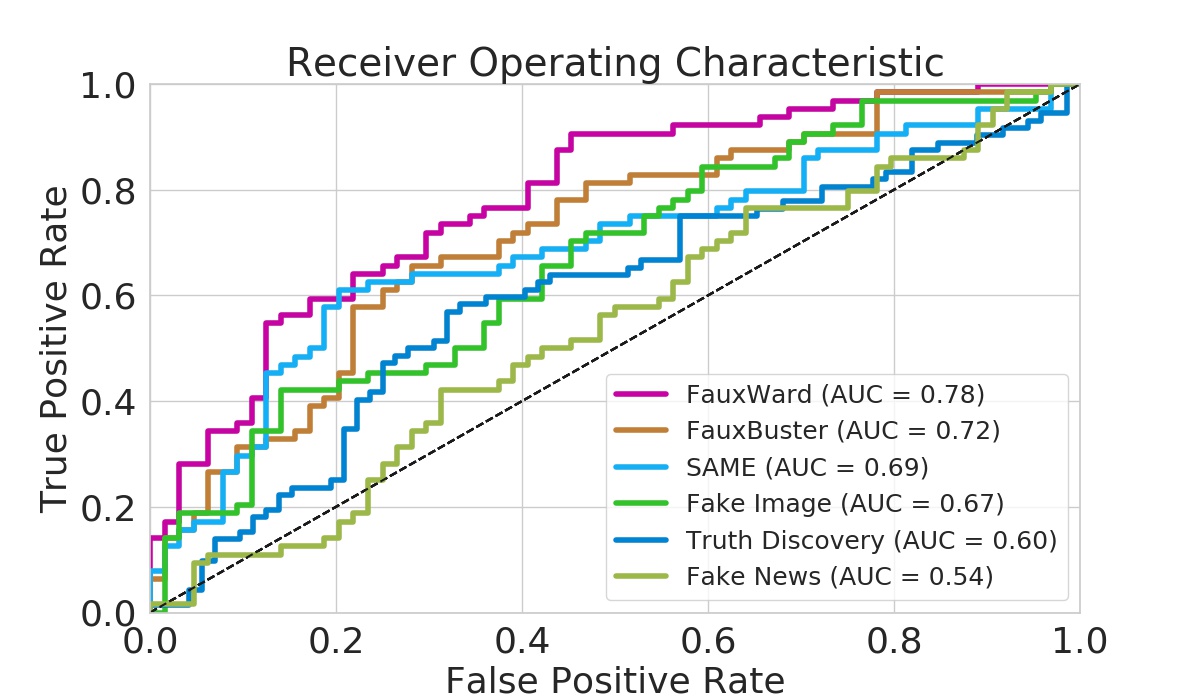}
    \caption{\shang{ROC Curve of All Schemes (Twitter)}}
    \label{fig:roct}
\end{figure}

\lanyu{We also plot the Receiver Operating Characteristics (ROC) curve of all methods in Figure \ref{fig:rocr} and \ref{fig:roct}. The ROC curve focuses on the trade-off between the False Positive Rate (FPR) and the True Positive Rate (TPR) by adjusting the classification threshold of each method. We observe that the FauxWard scheme continues to outperform all baselines in terms of the Area Under the Curve (AUC) score on both the Reddit and Twitter datasets. This demonstrates that FauxWard is also robust against the classification threshold.}

\subsection{FauxWard versus Humans}
\lanyu{In the second set of experiments, we compare the performance of FauxWard with humans.}
We invite three independent human annotators (denoted as A1, A2, and A3) to manually annotate whether they believe the image is misleading or not. We randomly pick a total of 70 image-based social media posts (38 of which are fauxtography) from the test dataset for them to annotate. Please note that these human annotators are different from the ground-truth annotators in that they have not seen those posts before and are ~\emph{not allowed} to have access to any external data source (e.g., Google Search, fact-checking websites) to validate their annotations. 
Furthermore, the annotators were asked to skip the posts that they happen to know the ground truth. 

\lanyu{First, we asked these participants to annotate image-centric posts by only showing them the image and the text of a post, which contains the same information a user receives from the social media feed. Next, we asked the participants to annotate the same set of posts but also showed them the comments of each post.
We design such an experiment process to evaluate whether the user comments from social media would assist humans in identifying fauxtography posts. Table \ref{tab:human} shows the performance of each individual annotator and their aggregated results based on the majority voting (i.e., ``overall without comments" and ``overall with comments").
We observe that FauxWard consistently outperforms the human annotators even if they are allowed to access the comments from social media users.
A possible reason is that humans are often easily affected \shang{by} their subjectivity and emotions. For example, we found all of the three human annotators fail to identify a fauxtography that shows an injured koala was rescued from the Australian bushfire in 2020 (the fact is that the koala was rescued from another event in 2015). 
In addition, we also observe that human performance is boosted significantly when the user comments are available to the annotators. Such an observation verifies our assumption on the usefulness of user comments in detecting fauxtography posts. Moreover, we also observe that the fauxtography posts with real images are more likely to convince the human annotators to believe the content of the post.}
This again demonstrates that the fauxtography detection problem is more challenging than merely detecting ``fake images".

\begin{table}[htb!]
    \small 
    \centering
    \caption{FauxWard vs. Human Performance}
    \begin{tabular}{l c c c c}
    \toprule
    \midrule
    & Accuracy & F1 &FPR & FNR\\
    \cmidrule(l){1-5}
    \textbf{FauxWard} & \textbf{0.7571} & \textbf{0.7733} & \textbf{0.2500} & \textbf{0.2368 } \\
    \cmidrule(l){1-5}
    A1 without comments & 0.3714 & 0.2667 & 0.4193 &0.7949\\
    \cmidrule(l){1-5}
    A1 with comments &0.5857 &0.5915 &0.3548 &0.4615\\
    \cmidrule(l){1-5}
    A2 without comments &0.3571 &0.2105  &0.3871 &0.8462\\
    \cmidrule(l){1-5}
    A2 with comments &0.5714 &0.5588 &0.3637 &0.4864\\
    \cmidrule(l){1-5}
    A3 without comments &0.4143 &0.3051 &0.3548 &0.7692\\
    \cmidrule(l){1-5}
    A3 with comments &0.6286 &0.6176 &0.3030 &0.4324\\  
    \cmidrule(l){1-5}
    Overall without comments &0.3857 &0.2456 &0.3548 &0.8205\\
    \cmidrule(l){1-5}
    Overall with comments &0.6143 &0.6197 &0.3226 &0.4359\\  
    \midrule
    \toprule
    \end{tabular}
    
 * ``Overall" denotes the majority vote of the three annotators.
\label{tab:human}
\end{table}

\subsection{Detection Time}
\begin{figure}[!htb]
    \centering
    \includegraphics[width=0.8\linewidth]{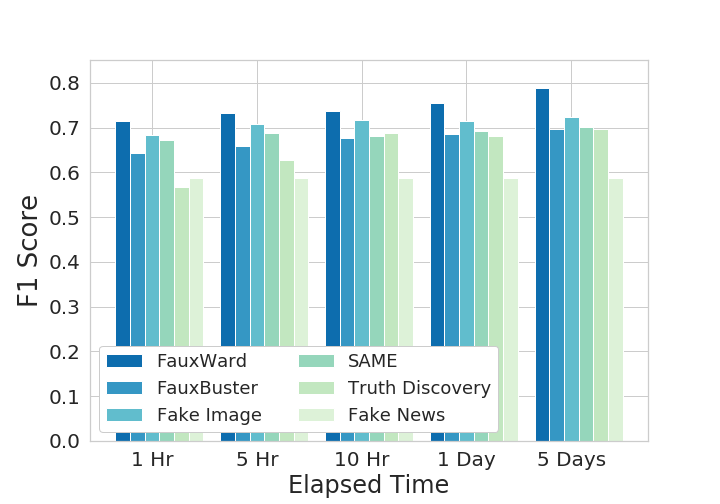}
    \caption{\shang{Elapsed Time vs. Performance (Reddit)}}
    \label{fig:timer}
\end{figure}

\begin{figure}[!htb]
    \centering
    \includegraphics[width=0.8\linewidth]{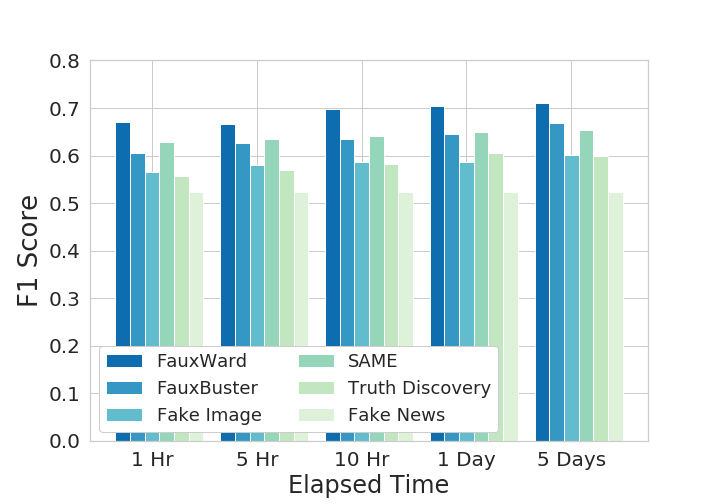}
    \caption{\shang{Elapsed Time vs. Performance (Twitter)}}
    \label{fig:timet}
\end{figure}

\lanyu{In the last set of experiments, we evaluate the detection performance of the FauxWard scheme against the time after a social media post is originally published. In particular, we limit the time window of the data collected from 1 hour to 5 days and only include user comments posted within the specific time window. The results are shown in Figure \ref{fig:timer} and Figure \ref{fig:timet}. We observe that the performance of the FauxWard scheme improves as the time increases, and more input data is available. In the meantime, FauxWard consistently outperforms all the baselines on both datasets. More importantly, FauxWard achieves a significant performance gain when the time window is short (e.g., 1 hour), which is necessary to curb the spread of misinformation on social media in a timely manner. 
}

\section{Conclusion}
\lanyu{In this paper, we develop a graph convolutional neural network approach, FauxWard, to address the fauxtography detection problem in image-based social media posts. FauxWard leverages the ``wisdom of the crowd" by exploring the valuable information from the user comments on social media and encoding the linguistic, sentiment, endorsement, and metadata attributes into a graph neural network framework. The FauxWard scheme does not directly analyze the content of image-centric posts and is robust against sophisticated content creators who are good at crafting and spreading the misleading fauxtography content on social media. We evaluate the FauxWard scheme using two real-world datasets collected from Reddit and Twitter. The results demonstrate that FauxWard can effectively detect the fauxtography posts on social media and outperforms the state-of-the-art baselines and human annotators in terms of accuracy and F1 score. }


\bibliographystyle{spbasic}
\bibliography{refs}

\end{document}